\documentclass[twocolumn,amsmath,floatfix,prb,aps]{revtex4}
\usepackage{color}
\usepackage{lipsum,amsmath}
\usepackage{pifont}   % Ding symbols
\usepackage{graphicx} % Include figure files
\usepackage{dcolumn}  % Align table columns on decimal point
\usepackage{bm}       % bold math
\usepackage{amsfonts} % Some more fonts
\usepackage{amssymb}  % More symbols
\usepackage{multirow} % Table functions
\usepackage{natbib}
\usepackage{siunitx}
\usepackage{supertabular}

\begin{document}
%\DeclareGraphicsRule{*}{png}{*}{}

%%%% User-defined commands %%%%
\newcommand{\ba}{{\bf a}}
\newcommand{\BB}{{\bf b}}
\newcommand{\bd}{{\bf d}}
\newcommand{\bD}{{\bf D}}
\newcommand{\br}{{\bf r}}
\newcommand{\bp}{{\bf p}}
\newcommand{\bk}{{\bf k}}
\newcommand{\bg}{{\bf g}}
\newcommand{\bt}{{\bf t}}
\newcommand{\bj}{{\bf j}}
\newcommand{\bu}{{\bf u}}
\newcommand{\bq}{{\bf q}}
\newcommand{\bG}{{\bf G}}
\newcommand{\bP}{{\bf P}}
\newcommand{\bJ}{{\bf J}}
\newcommand{\bK}{{\bf K}}
\newcommand{\bL}{{\bf L}}
\newcommand{\bR}{{\bf R}}
\newcommand{\bS}{{\bf S}}
\newcommand{\bT}{{\bf T}}
\newcommand{\bQ}{{\bf Q}}
\newcommand{\bA}{{\bf A}}
\newcommand{\bH}{{\bf H}}
\newcommand{\bff}{{\bf f}}

\newcommand{\bra}[1]{\left\langle #1 \right |}
\newcommand{\ket}[1]{\left| #1 \right\rangle}
\newcommand{\braket}[2]{\left\langle #1 | #2 \right\rangle}
\newcommand{\mel}[3]{\left\langle #1 \left| #2 \right| #3 \right\rangle}

\newcommand{\bdel}{\boldsymbol{\delta}}
\newcommand{\bsig}{\boldsymbol{\sigma}}
\newcommand{\beps}{\boldsymbol{\epsilon}}
\newcommand{\bnu}{\boldsymbol{\nu}}
\newcommand{\bnab}{\boldsymbol{\nabla}}
\newcommand{\bchi}{\boldsymbol{\chi}}
\newcommand{\bGam}{\boldsymbol{\Gamma}}
\newcommand{\bsigz}{\boldsymbol{\sigma_z}}

\newcommand{\bgt}{\tilde{\bf g}}

\newcommand{\brh}{\hat{\bf r}}
\newcommand{\bph}{\hat{\bf p}}

\author{R. Gupta$^1$}
\author{F. Rost$^1$}
\author{M. Fleischmann$^1$}
\author{S. Sharma$^2$}
\author{S. Shallcross$^1$}
\email{sam.shallcross@fau.de}
\affiliation{1 Lehrstuhl f\"ur Theoretische Festk\"orperphysik, Staudtstr. 7-B2, 91058 Erlangen, Germany,}
\affiliation{2 Max-Planck-Institut fur Mikrostrukturphysik Weinberg 2, D-06120 Halle, Germany.}

\title{Straintronics beyond homogeneous deformation}

\date{\today}

\begin{abstract}
We present a continuum theory of graphene treating on an equal footing both homogeneous Cauchy-Born (CB) deformation, as well as the microscopic degrees of freedom associated with the two sublattices. While our theory recovers all extant results from homogeneous continuum theory, the Dirac-Weyl equation is found to be augmented by new pseudo-gauge and chiral fields fundamentally different from those that result from homogeneous deformation. We elucidate three striking electronic consequences: (i) non-CB deformations allow for the transport of valley polarized charge over arbitrarily long distances e.g. along a designed ridge; (ii) the triaxial deformations required to generate an approximately uniform magnetic field are unnecessary with non-CB deformation; and finally (iii) the vanishing of the effects of a one dimensional corrugation seen in \emph{ab-initio} calculation upon lattice relaxation are explained as a compensation of CB and non-CB  deformation.
\end{abstract}

\maketitle
\section{Introduction}

With the emergence of two dimensional materials tantalizing new possibilities now exist to control electronic structure via material deformation \cite{rol15,AMORIM20161,shall17}. The most studied such system is graphene - an atomically thin layer of carbon - in which strain creates pseudo-magnetic and electric fields
\cite{2002-paper, ref13, PhysRevLett.108.227205, non-uniform-strain, AMORIM20161,Peeters-revisited}.
This connection between deformation and electromagnetic fields represents a far more profound control of electronic structure through deformation than is possible with any three dimensional material, and has led to a wealth of ideas for manipulating electronic currents in graphene via strain, together known as the field of ``straintronics''\cite{Valley-filter,Valley-filter2,Valley-filter3,Valley-filter4,Valley-filter5,Valley-filter6,Valley-filter7,Valley-filter8,Valley-filter9,Designing}.

In two dimensional materials lattice deformations often occur over length scales far in excess of the lattice constant, implying a natural role for a continuum description. For graphene this leads to a physically transparent theory connecting lattice deformation, via a pseudo-gauge in the Dirac-Weyl equation, to the remarkable experimental finding of Landau levels in the absence of an external magnetic field\cite{Klimov1557,Science-NB,Quantized,PhysRevB.87.205405,Landau-quantization2,77K-ridge}. A common assumption of such theories, however, is the Cauchy-Born rule \cite{CB-Rule} that states deformations around any material point are homogeneous. For a material with a lattice and basis, this implies the basis atoms of the unit cell deform according to a single global deformation field: there are no internal degrees of freedom between the sublattices. However, there is growing evidence from atomic simulations\cite{midgap,Carbon,Interplay,nanobubbles1,Peeters1,Peeters2,Peeters-triaxial,MD-study,midgap2,Relaxation} that in graphene, as for other  carbon allotropes such as diamond, this assumption breaks down.

What is therefore required is a continuum approach beyond the Cauchy-Born approximation: one that bridges the micro- and meso-scales of deformation.
The purpose of the present paper is to describe such an approach.
To that end, we augment the homogeneous ``acoustic'' deformation field with an ``optical'' field describing sub-lattice internal degrees of freedom, and develop an electronic theory that treats these two equally. While our focus is graphene, the framework we describe is easily generalized to any non-Bravais material, and we indicate how this may be done. 

We show that the electronic manifestation of deformations beyond the Cauchy-Born rule can be dramatic. In particular we find that non-Cauchy-Born deformations: (i) can create approximately uniform magnetic fields without recourse to special triaxial deformations\cite{Guinea2009}; (ii) allow the possibility of valley polarized charge transport over extended distances e.g. along a designed ridge; and (iii) qualitatively change patterns of charge localization and associated sub-lattice polarization. These features all arise as the introduction of a non-Cauchy-Born component profoundly changes the functional relationship between deformation field and pseudo-gauge. In contrast to the fundamental ``entanglement'' of the lattice geometry with the pseudo-gauge in homogeneous deformation\cite{multi-uni,PhysRevB.93.035456,Gomes2012,SSB4,SSB5,SSB3,currents2,PhysRevB.90.041411,currents3,Local,Electronic,Valley-filter}, in the non-Cauchy-Born case the pseudo-gauge depends only on the deformation field itself. So, for example, the nodal ($B=0$) lines of the resulting pseudo-magnetic fields reflect basic structures of the deformation field (e.g. a change in sign of its curvature) rather than the $C_3$ symmetry of the underlying lattice, allowing transport of charge along snake states associated with extended nodal lines, for instance created by a ridge or step edge.

We also examine atomic simulations of deformation in graphene that exist in the literature, and argue that a number of unexpected findings, in particular the vanishing of gauge field effects upon relaxation of armchair corrugation deformations\cite{midgap,Carbon}, yield to simple explanation in the generalized continuum theory presented here. Taken together, these results show that the possibilities of ``straintronics'' in graphene can be profoundly enriched by the inclusion of deformations beyond the Cauchy-Born rule.
 
%%%%%%%%%%
% Theory %
%%%%%%%%%%

\section{Theory}

We consider two distinct deformation fields $\bu^{(\nu)}(\br) = (u^{(\nu)}_x(\br), u^{(\nu)}_y(\br), u^{(\nu)}_z(\br))$ $\nu = 1,2$, each applied to one of the two sub-lattices of graphene. Here $\br$ is a 2-vector describing a position in the material, and the $\bu^{(\nu)}$ a 3-vector, allowing for both out-of-plane and in-plane deformations. In general, there will be $S$ such fields for a material with $S$ sub-lattices. While the formalism we describe here can be easily generalized, for the purposes of clear exposition (and as we only consider graphene) we will restrict to $S=2$. These two fields are conveniently expressed as 

\begin{equation}
\bu^\pm(\br) = \frac{1}{2}(\bu^{(1)}(\br)\pm\bu^{(2)}(\br))
\label{optdef}
\end{equation}
with $\bu^+(\br)$ the acoustic field that describes homogeneous Cauchy-Born obeying deformations, and $\bu^-(\br)$ an optical field encoding the internal degree of freedom between basis atoms in the unit cell. (In the following we will use the terms Cauchy-Born and acoustic, and non-Cauchy-Born and optical interchangeably.) Having established the form of deformation that we will consider we then review in Section \ref{OPEQUIV} an \emph{exact} mapping of the Slater-Koster tight-binding Hamiltonian to a general continuum operator $H(\br,\bp)$\cite{M}. This exact map is then in Section \ref{EX} applied to our specific deformation and, by Taylor expanding for slowly varying fields and small momenta, we recover a systematic series of contributions to the effective Hamiltonian that describe with increasing accuracy CB and non-CB deformations in graphene. 

In Section \ref{TAB} we describe in detail the resulting effective Hamiltonian. We recover all terms found in the standard continuum theory of deformations in graphene, including the gauge and scalar field terms\cite{Electronic,2002-paper,ref13, PhysRevLett.108.227205,non-uniform-strain, AMORIM20161,Peeters-revisited}, their curvature corrections\cite{PhysRevLett.108.227205}, as well as the Fermi velocity\cite{non-uniform-strain, AMORIM20161,Peeters-revisited,JANG2014139} and cone tilting expressions\cite{AMORIM20161}. In addition, we are also able to reproduce the limited number of non-CB results already found in the literature\cite{non-uniform,lin12,ref13}. However, the approach here in which CB and non-CB deformations are treated on an equal footing leads, as we show, to a wealth of new structures in the effective Hamiltonian.

\subsection{The ``operator equivalent'' approach} \label{OPEQUIV}

We first describe in outline the exact map from an atomistic tight-binding (TB) Hamiltonian $H_{TB}$ to a general continuum operator $H(\br,\bp)$. A complete  derivation of the formalism presented here can be found in Ref.~\onlinecite{M}, where it is also generalized to deal with multilayer situations. It has previously been applied to study interlayer deformations in bilayer graphene, treating both dislocations\cite{kiss15,shall17} as well as twist and shear faults\cite{vogl16,ray16}; this work represents the first application to study single layer deformations.

We consider a general TB Hamiltonian 

\begin{equation}
 H_{TB} = \sum_{ij} t_{ij} c_j^\dagger c_i
\end{equation}
where $t_{ij}$ are the overlap integrals of the crystal potential with local orbitals, with $c_i$ ($c_i^\dagger$) the annihilation and creation operators of these local orbitals. Note that we employ here a compressed index notation which minimally represents a basis atom of the unit cell (as is the case for graphene in the $\pi$-band approximation), but more generally can include spin and angular momentum labels.

We wish to obtain an effective continuum Hamiltonian that is \emph{exactly} equivalent to the TB Hamiltonian. The general approach, as described in Ref.~\onlinecite{M}, is firstly to break down the TB Hamiltonian into high and low symmetry parts:

\begin{eqnarray}
 H_{TB} & = & \sum_{ij} t^{(0)}_{ij} c_j^\dagger c_i + \sum_{ij} \delta t_{ij} c_j^\dagger c_i \\
  & = & H_{TB}^{(HS)} + H_{TB}^{(LS)}
\end{eqnarray}
where $t^{(0)}_{ij}$ are the overlap integrals of a high-symmetry state, and $\delta t_{ij}$ the changes induced by some deformation applied to it. Note that $H_{TB}^{(HS)}$ and $H_{TB}^{(LS)}$ share the same basis of local orbitals, differing only in the values of the hopping constants.

For operator equivalence we require two conditions: (i) a one to one correspondence between the complete basis sets of each Hamiltonian $H_{TB}$ and $H(\hat{\br},\hat{\bp})$ and (ii) equality between all inner products that can be constructed with each Hamiltonian and its basis set. For the TB Hamiltonian the choice of basis set is the Bloch states of the high symmetry part of the Hamiltonian $H_{TB}^{(HS)}$: 

\begin{equation}
 \ket{\Psi_{\bk\alpha}} = \frac{1}{\sqrt{N}}\sum_i e^{i\bk.(\bR_i + \bnu_\alpha)}\ket{\bR_i + \bnu_\alpha}
 \label{TBbas}
\end{equation}
with $\bR_i$ the lattice vectors and $\bnu_\alpha$ the basis vectors of the high symmetry structure, and $N$ the number of unit cells (the implicit $N\to\infty$ limit is suppressed, as is the $V\to\infty$ limit for the continuum representation below). The corresponding basis set for the continuum Hamiltonian are pseudospinor plane waves:

\begin{equation}
\label{Cbas}
 \ket{\phi_{\bk\alpha}} = \frac{1}{\sqrt{V}} e^{i\bk.\br} \ket{1_{\alpha}},
\end{equation}
where $\ket{1_{\alpha}}$ is a unit ket in a space with dimensionality equal the sum of atomic degrees of freedom,
i.e. $\ket{1_{\alpha}} = (0_{1},\ldots,1_{\alpha},\ldots)^T$, which will generally include other atomic degrees of freedom besides the basis index of the unit cell. These two sets of basis functions are in obvious one to one correspondence, as the number of atomic degrees of freedom $\alpha$ in the TB basis function, Eq.~\eqref{TBbas}, is equal to number of components of the pseudo-spinor vector in Eq.~\eqref{Cbas}. Having thus fulfilled the first of the two conditions described above, we can now precisely state the second:

\begin{equation}
 \mel{\Psi_{\bk_1\alpha}}{H_{TB}}{\Psi_{\bk_2\beta}} = \mel{\phi_{\bk_1\alpha}}{H(\br,\bp)}{\phi_{\bk_2\beta}}
 \label{opequiv}
\end{equation}
for all $\bk_1$, $\bk_2$, $\alpha$, and $\beta$. Surprisingly, as shown in Ref.~\onlinecite{M}, this condition can be exactly met and a closed form result for $H(\br,\bp)$ derived. The only  ingredient required to connect the atomistic and continuum worlds of Eq.~\eqref{opequiv} is to introduce an envelope function $t_{\alpha\beta}(\br,\bdel)$ that describes the overlap integral between an $\alpha$-orbital at a point $\br$ in the material to a $\beta$-orbital at point $\br+\bdel$. This function must evidently satisfy $t_{\alpha\beta}(\br,\bdel) = t_{ij}$ when $\br = \br_i$ and $\br+\bdel = \br_j$. The ``operator equivalent'' Hamiltonian is then found to be\cite{M}

\begin{equation}
 \left[H(\brh,\bph)\right]_{\alpha\beta} = \frac{1}{V_{UC}}\sum_{i}  M_{i\alpha\beta} 
 t_{\alpha\beta}(\br,\bK_i+\bp/\hbar)
 \label{Heff2}.
\end{equation}
In this expression the momentum operator $\bp$ is measured from some (arbitrary) point $\bK_1$ in the Brillouin zone of the high symmetry (HS) system, and the sum $i$ is taken over the translation group of the HS system, i.e. $\bK_i = \bK_1 + \bG_i$ with $\bG_i$ a reciprocal lattice vector. $M_{i\alpha\beta}$ is an element of a so-called ``M-matrix'', an object that encodes the geometry of the HS system, and is given by
\begin{equation}
 M_{i\alpha\beta} = e^{i(\bK_i-\bK_1).(\bnu_\alpha-\bnu_\beta)}.
 \label{Mmat}
\end{equation}
Finally $t_{\alpha\beta}(\br,\bq)$ is the mixed space hopping function, the Fourier transform with respect to $\bdel$ of the envelope function $t_{\alpha\beta}(\br,\bdel)$:
\begin{equation}
 t_{\alpha\beta}(\br,\bq) = \int d\bdel e^{i\bq.\bdel} t_{\alpha\beta}(\br,\bdel).
\end{equation}
For further details of the derivation we refer the reader to Ref.~\onlinecite{M}, however we mention here one subtle detail. The operator equivalence in Eq.~\eqref{opequiv} is posited on the deformation changing the Hamiltonian while the basis is held fixed. As the basis set is complete for any deformation this is allowed (since local orbitals are neither created nor destroyed, i.e. the number of sites in the crystal remains unchanged under deformation). However, a fixed basis means also fixed labels of the basis functions $\ket{\bR_i + \bnu_\alpha}$, and unchanging position labels under deformation imply in turn a coordinate system co-moving with the deformation. Thus, both $\br$ and $\bp$ are measured in a local coordinate system, explaining the presence in Eqs.~\eqref{Heff2} and \eqref{Mmat} of the reciprocal lattice quantities of the high symmetry system.

\subsection{Derivation of the effective Hamiltonian for pristine graphene} \label{PRIS}

As a simple example we first derive the Hamiltonian for pristine graphene using the above formalism. Using the high symmetry K point as the reference point for measuring momenta $\bp$, the $C_3$ symmetry then gives a first star of three vectors: $K_0 = \left(\frac{2}{3},0\right)$ and $K_\pm = \left(-\frac{1}{2},\pm\frac{1}{2\sqrt{3}}\right)$, with corresponding $\bG_i$ given by $\bG_0 = (0,0)$, $\bG_\pm = \left(-1,\pm\frac{1}{\sqrt{3}}\right)$. For a choice of basis vectors $\bnu_1 = (0,0)$ and $\bnu_2 = \left(\frac{1}{2},\frac{1}{2\sqrt{3}}\right)$ the corresponding ``M matrices'' are given by:

\begin{equation}
 M_0 = \begin{pmatrix} 1 & 1 \\ 1 & 1 \end{pmatrix},~~~ M_\pm = \begin{pmatrix} 1 & e^{\pm i 2\pi/3} \\ e^{\mp i 2\pi/3} & 1\end{pmatrix}
\end{equation}
(note that reciprocal space quantities are in units of $2\pi/a$, real space quantities in units of $a$).
Evaluation of the effective Hamiltonian now only requires a choice of hopping function. For pristine graphene the low energy $\pi$-band can be completely described by $t_{pp\pi}$ tight-binding integrals, and so the hopping function has no labels and is given simply by $t(\bdel^2)$. In the case of out-of-plane deformation a full treatment of the angular degrees of freedom of the hopping vector requires inclusion of both $t_{pp\sigma}$ and $t_{pp\pi}$ hopping via the usual Slater-Koster scheme, evidently possible with the approach of Section \ref{OPEQUIV} and which we will describe later. 

Taylor expansion about the high symmetry K point then yields

\begin{equation}
 H(\bp) = \sum_n T_n \left(\frac{p}{\hbar}\right)^n
 \label{HHS}
\end{equation}
where $n$ is a tuple of two integers $(n_1, n_2)$ and $(p/\hbar)^n = (p_x/\hbar)^{n_1} (p_y/\hbar)^{n_2}$. We will employ this shorthand multi-index tuple notation throughout this paper. In Eq.~\eqref{HHS} $T_n = \frac{1}{n! V_{UC}} \sum_i M_i \left.\partial_q^n t(q^2)\right|_{q=K_i}$ and so carries the $SU(2)$ matrix structure ($n! = n_1!n_2!$). It easily evaluated in the first star approximation described above to yield, for zeroth and linear order, $T_{00} = E_0 \sigma_0$, $T_{10} = v_F \sigma_x$, $T_{01} = v_F \sigma_y$. Here the constants depend on $V_{UC}$ and $t(\bK_i^2)$ with $K_i$ the translation group of the high symmetry K point and $t(\bq^2)$ the Fourier transform of the tight-binding hopping function. In this way we find for pristine graphene up to second order in momentum:

\begin{equation}
 H(\bp) = v_F \bsig.\bp + \frac{p^2}{2m} \sigma_0 + \beta \begin{pmatrix} 0 & (p_x + i p_y)^2 \\ (p_x - i p_y)^2 & 0 \end{pmatrix}
\end{equation}

\subsection{Derivation of the effective Hamiltonian for deformed graphene} \label{EX}

We now implement deformation within this general effective Hamiltonian scheme. Firstly we expand the mixed space hopping function in Eq.~\eqref{Heff2} close to the Dirac point. The expression $t_{\alpha\beta}(\br,\bK_i+{\bp}/\hbar)$ then becomes

\begin{equation}
 t_{\alpha \beta}(\br,\bK_i+\bp/\hbar) = \sum_{n} \frac{1}{n!} \partial_{q}^{n}\left.t_{\alpha\beta}(\br,\bq)\right|_{\bq=\bK_i}\left(\frac{{p}}{\hbar}\right)^{n}
 \label{t_final}
\end{equation}

The next step is to obtain the mixed space hopping function $t_{\alpha\beta}(\br,\bq)$. For clarity of exposition we will employ here a $\pi$-orbital only (H\"uckel model) approximation for deformed graphene; this restriction will subsequently be removed when we consider a more general scheme involving the full Slater-Koster expression for the $\pi$-band including both $\sigma$-hopping ($t_{pp\sigma}$) and $\pi$-hopping ($t_{pp\pi}$), with the $\sigma$-hopping resulting from the bending of $p_z$ orbitals of graphene under out-of-plane deformation. 

\begin{table*}[]
\begin{tabular}{|c|c|c|c|c|c|}
\hline
Order ($|m|$) &  $m_1 m_2$ &$C_{1m}$ & $C_{2m}$ & $C_{3m}$ & $C_{4m}$ \\
\hline
\hline
0 & 0 0 & 0 & 0 & $4(\bu^-)^2$ &0\\
\hline
1 & 1 0  & 0 & 0 &$4\bu^-\cdot\partial_x\bu^-$ & $4u_x^- + 2 \bu^- . \partial_x  \bu^+$\\
& 0 1 & 0 & 0 &$4\bu^-\cdot\partial_y\bu^-$ & $4u_y^- + 2 \bu^- . \partial_y  \bu^+$\\
\hline
& 2 0  & $2 \epsilon_{xx}^+$ & $2 (\epsilon_{xx}^- +\partial_x \bu^+ . \partial_x  \bu^-) $ & $2\bu^-.\partial_x^2\bu^-$ & $2 \epsilon_{xx}^- $\\
2 & 1 1 & $4\epsilon_{xy}^+ $ & $2(2\epsilon_{xy}^- + \partial_x \bu^+ . \partial_y \bu^- + \partial_y \bu^+ . \partial_x \bu^-)$ & $4\partial_x\bu^-.\partial_y\bu^-$ & $4\epsilon_{xy}^-$\\
 & 0 2 & $2 \epsilon_{yy}^+ $ & $2 (\epsilon_{yy}^- +\partial_y \bu^+ . \partial_y  \bu^-)$ & $2\bu^-.\partial_y^2\bu^-$ & $2 \epsilon_{yy}^-$\\
\hline
& 3 0 & $\partial_x \epsilon_{xx}^+$ & $\partial_x \epsilon_{xx}^-$ & - & $\partial_x \epsilon_{xx}^-$ \\
3 & 2 1 & $\partial_x (\epsilon_{xy}^++2\epsilon_{yx}^+$) & $\partial_x (\epsilon_{xy}^-+2\epsilon_{yx}^-$) & - & $\partial_x (\epsilon_{xy}^-+2\epsilon_{yx}^-)$\\
& 1 2 & $\partial_y (2\epsilon_{xy}^++\epsilon_{yx}^+)$ & $\partial_y (2\epsilon_{xy}^-+\epsilon_{yx}^-)$ & - & $\partial_y (2\epsilon_{xy}^-+\epsilon_{yx}^-)$\\
& 0 3 & $\partial_y \epsilon_{yy}^+$ & $\partial_y \epsilon_{yy}^-$ & - & $\partial_y \epsilon_{yy}^-$\\
\hline
\end{tabular}
\caption{The expansion coefficients $C_{\eta m}$ of the function describing the change in electron hopping due to deformation $t_{\eta}(\br,\bdel)=t^{(1)}(\delta^2)\sum_{m}{C_{\eta m}}(\br)\delta^m$. Here $\eta = 1$ corresponds to Cauchy-Born deformation, and $\eta > 1$ to non-Cauchy-Born deformation (see Eq.~\eqref{XXX}). In each of the expressions shown $\beps^\pm$ and $\bu^\pm$ are the deformation field and deformation tensor for acoustic (+ superscript) and optical (- superscript) deformations. Only those coefficients consistent with the hermiticity conditions Eq.~\eqref{C1} and Eq.~\eqref{C2} are shown, while those that violate the hermiticity of the effective Hamiltonian are removed as indicated by ``-'' in the $C_{3m}$ column.}\label{t-table}
\end{table*}

As the only impact of deformation is to change the values of the hopping integrals (the basis being held fixed), to obtain $t_{\alpha\beta}(\br,\bdel)$ we require only (i) a form of the electron hopping envelope function $t(\bdel)$ in the high symmetry system, and (ii) information about how an arbitrary hopping vector $\bdel$ changes under deformation. As the origin $\br$ of the hopping vector is on the $\alpha$-sublattice, and the end point $\br+\bdel$ on the $\beta$-sublattice, and as we have different deformation fields acting on each of these sublattices, then the change under deformation is: $\bdel \to \bdel+\bu^{(\beta)}(\br+\bdel)-\bu^{(\alpha)}(\br)$. Employing the same form of hopping function for pristine graphene as used in the previous section, we find that under deformation it changes as
\begin{equation}
 t(\bdel^2) \to t_{\alpha\beta} \left((\bdel+\bu^{(\beta)}(\br+\bdel)-\bu^{(\alpha)}(\br))^2\right).
 \label{tdef}
\end{equation}
To obtain a complete picture of the hopping within the unit cell it is convenient to write the hopping function as a $2\times2$ matrix in sublattice space. Introducing the optical and acoustic fields $\bu^{\pm}(\br)$, Eq.\eqref{optdef}, we find
\begin{widetext}
\begin{equation}
\label{tmat}
 \bt(\br,\bdel) = \begin{pmatrix}
      t\left((\bdel + D(\bu^++\bu^-))^2\right) &
      t\left((\bdel - 2\bu^- + D(\bu^+-\bu^-))^2\right) \\
      t\left((\bdel + 2\bu^- + D(\bu^++\bu^-))^2\right) &
      t\left((\bdel + D(\bu^+-\bu^-))^2\right)      
     \end{pmatrix},
\end{equation}
\end{widetext}
where the operator shorthand $D\bff = \bff(\br+\bdel)-\bff(\br)$ has been introduced. Performing a first order Taylor expansion of the function $t$, the change of the hopping function in sub-lattice space is found to be
\begin{eqnarray}
 \delta {\bt}(\br,\bdel) & = & \begin{pmatrix} 1 & 1 \\ 1 & 1 \end{pmatrix} t_1(\br,\bdel)
 + \begin{pmatrix} 1 & 0 \\ 0 & 1 \end{pmatrix}\sigma_z t_2(\br,\bdel) \nonumber\\
 & + & \begin{pmatrix} 0 & 1 \\ 1 & 0 \end{pmatrix} t_{3}(\br,\bdel) +
 \begin{pmatrix} 0 & 1 \\ 1 & 0 \end{pmatrix}\sigma_z t_{4}(\br,\bdel)
\label{XXX}
\end{eqnarray}
where

\begin{eqnarray}
\label{XX}
 t_1(\br,\bdel) & = & 2t^{(1)}(\bdel^2)\left(\bdel\cdot D\bu^+ + (D\bu^+)^2 + (D\bu^-)^2\right), \nonumber\\
 t_2(\br,\bdel) & = & 2t^{(1)}(\bdel^2)(\bdel\cdot D\bu^- + D\bu^+\cdot D\bu^-), \nonumber\\
 t_3(\br,\bdel) & = & 4t^{(1)}(\bdel^2)\left((\bu^-)^2 + D\bu^-\cdot\bu^-\right), \nonumber\\
 t_4(\br,\bdel) & = & 2t^{(1)}(\bdel^2)\bigl(2\bdel\cdot\bu^- + \bdel\cdot D\bu^-+2D\bu^+\cdot \bu^- \nonumber\\
&&+D\bu^+\cdot D\bu^-\bigr)
\end{eqnarray}
with $t^{(1)}(\bdel^2)=\partial t(\bdel^2)/\partial(\bdel^2)$. These four linearly independent matrices encapsulate Cauchy-Born law and beyond Chaucy-Born law deformations in graphene. The first matrix describes a hopping change homogeneous in sub-lattice space, this evidently represents the CB obeying part of the deformation. The remaining matrices encode inhomogeneous hopping in sub-lattice space. At zeroth order in momentum the second of these matrices represents chiral (mass generating) fields due to non-CB deformation (note the presence of $\sigma_z$), while the remaining two represent new gauge fields. At higher order in momentum all three will generate velocity and trigonal warping corrections to the effective Hamiltonian.

To make further progress we now perform a Taylor expansion of the $D\bff$ type terms in Eq.~\eqref{XX}. For any of the $t_\eta$ ($\eta = 1-4$) the result may evidently be expressed as

\begin{equation}
\label{t_r}
t_{\eta}(\br,\bdel)=t^{(1)}(\delta^2)\sum_{m} {C_{\eta m}}(\br)\delta^m
\end{equation}
where $C_{\eta m}$ are the coefficients that depend on the deformation field $\bu^\pm(\br)$ and $m=(m_1,m_2)$ is a tuple of integers that correspond to the power of $\delta_x$ and $\delta_y$, respectively. For the four hopping functions in Eq.~\eqref{XX} the $C_{\eta m}$ coefficients are shown in Table \ref{t-table}, where we restrict ourselves to those $C_{\eta m}$ that will ultimately preserve hermiticity of the effective Hamiltonian (see in Section \ref{Sec-Herm}). The coefficients of the $t_1(\br,\bdel)$ expansion are the familiar coefficients of a bond deformation in a single deformation field $u^+(\br)$, while the coefficients for $t_2(\br,\bdel)$-$t_4(\br,\bdel)$ reflect the action of two distinct deformation fields on the two end points of the bond.

The Fourier transform with respect to $\bdel$ is now trivial and gives
\begin{equation}
\label{t_q}
t_{\eta}(\br,\bq)=\sum_m(-i)^m C_{\eta m}(\br)\partial_q^m t^{(1)}(q^2)
\end{equation}
where

\begin{equation}
t^{(1)}(q^2) = \int d\bdel e^{i\bq.\bdel} t^{(1)}(\delta^2).
\end{equation}
Denoting the matrix corresponding to $t_\eta(\br,\bdel)$ in Eq.~\eqref{XXX} as $L_\eta$, and inserting both into Eq.~(\ref{t_final}) we find the expression

\begin{eqnarray}
\label{TRP}
\delta t_{\alpha\beta}(\br,\bK_i+\bp/\hbar) & = &  \sum_{\eta} L_{\eta\alpha\beta} \sum_{nm}\frac{(-i)^m}{n!} \\
& \times & C_{\eta m}(\br)\partial_q^{m+n} t^{(1)}(q^2)|_{q=K_i}\left(\frac{p}{\hbar}\right)^n\nonumber
\end{eqnarray}
which can now be inserted back into Eq.~\eqref{Heff2} to arrive at a compact expression for the effective Hamiltonian of graphene with both Cauchy-Born and non-Cauchy-Born deformations

\begin{equation}
\label{deformed}
\delta H(\br,\bp) =\sum_{\eta n m}\frac{1}{n!} C_{\eta m}(\br)T_{\eta,m+n}\left(\frac{p}{\hbar}\right)^n
\end{equation}
with

\begin{equation}
T_{\eta, m,\alpha\beta}=\frac{L_{\eta\alpha\beta}}{V_{UC}} \sum_i M_{i\alpha\beta}\partial_q^{m} t^{(1)}(q^2)|_{q=K_i}.
\label{Tm}
\end{equation}
$T_{\eta, m}$ is independent of position and momentum and, as in the example of pristine graphene, simply carries the matrix structure of the Hamiltonian. The position, momentum, and matrix degrees of freedom of the effective Hamiltonian thus factorize. While for graphene the position functions $C_{\eta m}(\br)$ are obviously purely geometric in origin, in a more complex material, or by going beyond $\pi$-band hopping, they will encode both the geometry of the deformed bond as well as the form of the Slater-Koster hopping function.

The generalization to $S$ deformation fields and a general Slater-Koster form of the hopping function proceeds straightforwardly, with no formal change in structure of the preceding equations but with Eq.~\ref{XXX} generalized to

\begin{equation}
 \delta\bt(\br,\bdel) = \sum_\eta L_\eta \,t_\eta(\br,\bdel),
\end{equation}
where $\eta$ is now a generalized index that counts deformation modes as well as the spherical and cylindrical angular momenta of Slater-Koster integrals, i.e. $t_{ss\sigma}$, $t_{sp\sigma}$, $t_{pp\sigma}$, $t_{pp\pi}$ and so on.
For all $S$ and Slater-Koster forms this equation can be written down, but in contrast to the case of graphene the matrices $L_\eta$ are not guaranteed to form a linearly independent set.

\subsection{Scalar, gauge, and chiral fields due to Cauchy-Born and non-Cauchy-Born deformation} \label{TAB}

\begin{table*}[!t]
\begin{tabular}{|c|c|c|c|c|c|}
\hline
 & $i$& $\alpha_i$ &  $h_i$  & $\eta$ & Ref. \\
\hline
\hline
&&& Acoustic &&\\
\hline
\hline
&&&&&\\
 & 1 & $0.9$ & $\bigl[\bnab . \bu_\perp^++\bigl((\partial_x \bu^+)^2+(\partial_y\bu^+)^2\bigr)/2\bigr]\sigma_0$& 1 & \onlinecite{PhysRevLett.108.227205, non-uniform-strain, AMORIM20161} \\
& 2 & $8.5$ & $\bigl[\{\epsilon_{xx}^+-\epsilon_{yy}^+ + \bigl((\partial_x \bu^+)^2-(\partial_y\bu^+)^2\bigr)/2\}\sigma_x-(2\epsilon_{xy}^++ \partial_x \bu^+ . \partial_y \bu^+)\sigma_y\bigr]$& 1 & \onlinecite{2002-paper, ref13, PhysRevLett.108.227205, non-uniform-strain, AMORIM20161,Peeters-revisited}\\
Field & 3 & $i0.9$ & $[\epsilon_{xxx}^+ -(2\epsilon_{xyy}^+ + \epsilon_{yyx}^+)]\sigma_0$ & 1 & \\
& 4 & $-i2.9$ & $\begin{pmatrix} \sigma_x & \sigma_y
\end{pmatrix}
\begin{pmatrix}
3\epsilon_{xxx}^+  + 2\epsilon_{xyy}^+ + \epsilon_{yyx}^+ \\
\epsilon_{xxy}^+ +2\epsilon_{yxx}^+ + 3\epsilon_{yyy}^+
\end{pmatrix}$ & 1 & \onlinecite{PhysRevLett.108.227205}\\
&&&&&\\
\hline
&&&&&\\
& 5 & $-1.8$ & $[(\epsilon_{xx}^+-\epsilon_{yy}^+)p_x-2\epsilon_{xy}^+)p_y]\sigma_0$& 1 & \onlinecite{AMORIM20161}\\
$\substack{ \mathrm{Fermi} \\\mathrm{Velocity}} $ & 6 & $5.6$ & $\begin{pmatrix} \sigma_x & -\sigma_y \end{pmatrix} \begin{pmatrix} 3\epsilon_{xx}^++\epsilon_{yy}^+ & 2\epsilon_{xy}^+ 
\\
2\epsilon_{xy}^+ & \epsilon_{xx}^++3\epsilon_{yy}^+
\end{pmatrix}\begin{pmatrix} p_x \\ p_y \end{pmatrix}$& 1 & \onlinecite{PhysRevLett.108.227205, non-uniform-strain, AMORIM20161,Peeters-revisited}\\
&&&&&\\
\hline
\hline
& &  & Optical &  & \\
\hline
\hline
&&&&&\\
& 7 &  $0.9$ & $\bnab . \bu_\perp^-\sigma_z + [((\partial_x \bu^-)^2 + (\partial_y \bu^-)^2)/2]\sigma_0$ & 2,1 &\\ 
 &  8 & $23.7$ & ${\boldsymbol{\sigma}\cdot(u_y^-,-u_x^-)}$ & 4 &\onlinecite{lin12, PhysRevB.76.045430}\\
 & 9 & $8.5$ & $\Bigl[\bu^-.\Bigl(\sigma_x(\partial_x^2-\partial_y^2)-2\sigma_y\partial_x\partial_y\Bigl)\bu^-\Bigr]$ & 3 &\\
 Field & 10 & $-i8.5$ & ${\bsig\cdot\Bigl[\Bigl(2\epsilon_{xy}^- + i\bigl[(\partial_x \bu^-)^2 - (\partial_y \bu^-)^2\bigr]/2\Bigr),\Bigl(\epsilon_{xx}^--\epsilon_{yy}^- - i \partial_x \bu^- . \partial_y \bu^-\Bigr)\Bigr]}$&4,1 & \\ & 11 & $-i23.7$ & 
$\bu^- . [\partial_x \bu^- \sigma_x +  \partial_y \bu^- \sigma_y]$
& 3 &  \\
 & 12 & $i0.9$ & $[\epsilon_{xxx}^- -(2\epsilon_{xyy}^- + \epsilon_{yyx}^-]\sigma_z$ & 2 &\\
& 13 & $2.9$ & $\begin{pmatrix} -\sigma_y & \sigma_x
\end{pmatrix}
\begin{pmatrix}
3\epsilon_{xxx}^-  + 2\epsilon_{xyy}^- +  \epsilon_{yyx}^- \\
\epsilon_{xxy}^- + 2\epsilon_{yxx}^- + 3\epsilon_{yyy}^-
\end{pmatrix}$ & 4 &\\
&&&&&\\
\hline
&&&&&\\
& 14 & $-1.8$ & $\Bigl[(\epsilon_{xx}^--\epsilon_{yy}^-)p_x-(2\epsilon_{xy}^-)p_y\Bigr]\sigma_z $ & 2 & \\
 & 15 & $17.1$ & $ \begin{pmatrix} \sigma_x & \sigma_y \end{pmatrix}
\begin{bmatrix} 
{u_y^-} & {u_x^-} \\
{u_x^-} & {-u_y^-}
\end{bmatrix}
\begin{pmatrix}
p_x \\ p_y
\end{pmatrix}$ & 4 & \onlinecite{lin12,Optical-cond}\\
$\substack {\mathrm{Fermi}\\\mathrm{Velocity}}$ & 16 & $23.7$ &$(\bu^-)^2\boldsymbol (\sigma_x p_x + \sigma_y p_y)$ & 3 &\\
& 17 &$i5.9$ &$\begin{pmatrix} -\sigma_y & \sigma_x \end{pmatrix} \begin{pmatrix} 3\epsilon_{xx}^-+\epsilon_{yy}^- & 2\epsilon_{xy}^- 
\\
2\epsilon_{xy}^- & \epsilon_{xx}^-+3\epsilon_{yy}^-
\end{pmatrix}\begin{pmatrix} p_x \\ p_y \end{pmatrix}$ & 2 &\\
 & 18 &  $-i17.1$ & $ \bu^- .[(- \partial_x \bu^- \sigma_x + \partial_y \bu^- \sigma_y)p_x + ( \partial_y \bu^- \sigma_x+ \partial_x \bu^- \sigma_y) p_y] $
& 3 & \\
&&&&& \\
\hline
&&&&&\\
 & 19 & $-11.8$ & $\sigma_x u_y^-(p_x^2+3p_y^2)-\sigma_y u_x^- (3 p_x^2 + p_y^2)+(\sigma_x u_x^- - \sigma_y u_y^-)p_x p_y$ & 4 & \\
$\substack {\mathrm{Trigonal} \\ \mathrm{Warping}}$ & 20 & $-17.1$ & $({\bu^-})^2\Big[(p_x^2-p_y^2)\sigma_x-p_xp_y\sigma_y\Big]$ & 3 & \\
 &&&&&\\
\hline
\hline
& & &   Opto-acoustic & & \\
\hline
\hline
&&&&&\\
& 21 & $0.9$ & $(\partial_x \bu^- . \partial_x \bu^+ + \partial_y \bu^- . \partial_y \bu^+) \sigma_z$ & 2 &\\
 Field & 22 & $23.7$ & $(\bu^-\cdot\partial_y\bu^+)\sigma_x-(\bu^-\cdot\partial_x\bu^+)\sigma_y$ & 4 &\\
 & 23 & $-i 8.5$ & $\Big[\partial_x (\bu^-\cdot\partial_y\bu^+) +\partial_y (\bu^-\cdot\partial_x\bu^+)\Big]\sigma_x + \Big[\partial_x (\bu^-\cdot\partial_x\bu^+)-\partial_y (\bu^-\cdot\partial_y\bu^+)\Big]\sigma_y$ & 4 &\\
 &&&&&\\
 \hline
&&&&&\\
$\substack{\mathrm{Fermi}\\ \mathrm{Velocity}}$
& 24 & $17.1$ & $\left[(\bu^-\cdot\partial_x\bu^+)(\sigma_xp_y+\sigma_yp_x)+(\bu^-\cdot\partial_y\bu^+)(\sigma_xp_x-\sigma_yp_y)\right]$ & 4 &\\
&&&&&\\
\hline
\end{tabular}
\caption{
The effective Hamiltonian of graphene in the presence of homogeneous Cauchy-Born (acoustic) and non-Cauchy-Born (optical) deformation fields; the latter correspond to the internal degrees of freedom associated with the sublattices of graphene. $\epsilon^\pm_{ij}$ are components of the in-plane deformation tensors and $\bu^\pm$ the 3-vector deformation fields which therefore include both in-plane and out-of-plane components; $+$ denotes an acoustic field and $-$ an optical field. The effective Hamiltonian is given by $H = H_0 + \sum_i \alpha_i h_i$, with the coupling constant $\alpha_i$ given in the third column and the field expression $h_i$ in the fourth column. $H_0$ is the Hamiltonian of pristine graphene, see Eq.~\eqref{HHS}. Shown are effective field, velocity, and trigonal warping corrections for each deformation type i.e. acoustic, optical and their coupling (opto-acoustic). The fifth column $\eta$ is the hopping matrix type from which these terms have been derived (see Eq. \ref{t_r}), while terms already described in the literature have the corresponding bibliography reference displayed in the last column of the table.}
\label{Table_bful}
\end{table*}

We now describe the corrections to the Dirac-Weyl Hamiltonian of graphene that arise from Cauchy-Born (acoustic) and non-Cauchy-Born (optical) deformations, using the formalism derived in the previous section. For a first star approximation these may be obtained analytically from Eq.~\eqref{deformed} and \eqref{Tm}, but instead we have implemented these equations (generalized to $S$ deformation fields and underpinned by the full Slater-Koster tight-binding theory) into a software package for the general treatment of deformations in 2d materials. Thus all results are obtained via a numerical procedure with the ``star sums'' taken to numerical convergence. The resulting formulae are presented in Table \ref{Table_bful} and include terms up to second order in momentum and second order in spatial derivatives of the deformation fields. For ease of use in the following text the second column simply enumerates the various terms, which have been divided into those that arise from the acoustic field, the optical field, and their coupling (denoted opto-acoustic). Each entry displays the numerical coefficient of the expression $\alpha_i$ (column three), the expression itself $h_i$ (column four), the value of $\eta$ in Eq.~\eqref{XX} from which the term is derived (column five), and any references in which the expression has previously been reported (column six). The effective Hamiltonian due to deformation is then simply given by $\sum_i \alpha_i h_i$.

We first consider the effective fields that are generated from the $\eta=1$ term of the hopping function, Eq.~\eqref{XXX}. This term arises from homogeneous Cauchy-Born (i.e., acoustic) deformation of the lattice. These are displayed in terms 1-4 of Table II. 
These include the well known real (term 2) and imaginary (term 4) gauge fields, involving the deformation tensor and its derivative respectively. The corresponding real (term 1) scalar potential is also well known, however we also find an imaginary scalar potential (term 3) that does not, to the best of our knowledge, appear in the literature. This imaginary scalar potential is  the ``hermitian pair'' of the cone tilting expression (term 5) in the same way the imaginary gauge is the hermitian pair of the Fermi velocity correction (term 6), i.e. only when both these terms are included is the resulting Hamiltonian hermitian (the pairing of imaginary gauge and Fermi velocity was discussed in Ref.~\onlinecite{PhysRevLett.108.227205}). The question of hermiticity and hermitian pairs will be discussed carefully in the next section.

Examination of Eq.~\eqref{XX} reveals that the leading order term in the acoustic hopping function, $\bdel.D\bu^+$, reappears in two of the optical deformation hopping functions, $t_2$ and $t_4$ (with obviously $\bu^+$ replaced by $\bu^-$). These functions are multiplied by $\sigma_z$ in Eq.~\eqref{XXX}, leading to the following interesting correspondence rule: from each leading order acoustic term in the effective Hamiltonian an optical term can be obtained simply by multiplication with $\sigma_z$. This therefore sends scalar fields to chiral fields, and real gauge fields to imaginary gauge fields and vice versa according to $(A_x,A_y)\leftrightarrow i (-A_y,A_x)$. In this way terms $7, 10, 12$, and $13$ from $\eta=1$ can be directly obtained from the corresponding terms in $1, 2, 3$ and $4$, respectively.

While a constant acoustic deformation is simply a rigid shift of the lattice, without physical consequence, a constant optical deformation causes relative displacement of sublattices, obviously with physical consequence. This difference between acoustic and optical deformation implies terms in the Hamiltonian not covered by the correspondence rule above. This may be seen already in the coefficients $C_{3m}$ and $C_{4m}$ in Table~\ref{t-table}: the coefficients of $C_{4m}$ include terms (at $|m|=1$) that depend on the deformation field $\bu^-$ directly, while all of the $C_{3m}$ coefficients have no acoustic counterpart. These lead, respectively, to a zeroth order optical gauge term directly dependent on the optical deformation (term 8), previously been obtained in Refs.~[\onlinecite{lin12}] and [\onlinecite{PhysRevB.76.045430}], as well as gauge terms at higher order (terms 9, 11).

Finally, new effective fields arise from the coupling of the internal non-Cauchy-Born degrees of freedom to the homogeneous Cauchy-Born deformation field $\bu^+(\br)$. These are denoted opto-acoustic in Table II, and arise from the coefficients $C_{2m}$ and $C_{4m}$ in Table~\ref{t-table}. These produce a new chiral potential (term 21), as well as both real and imaginary gauge field (terms 22 and 23) respectively.

We note that there are also terms in Table II involving the square of the field (terms 1, 2, 7 and 10 in the table, for example). These terms are higher order corrections to the corresponding lower order terms, but are important for out-of-plane deformations which only enter at second order. This is a consequence of the mirror symmetry of pristine graphene in the $z$-direction, such that an out-of-plane deformation in the $+z$ or $-z$ direction is equivalent, and hence such deformations must involve the square of the deformation field.

We now consider the next order in the momentum expansion ($p=1$), where we find corrections to the Fermi velocity. Though the corrections arising from the acoustic field are well known in literature (terms 5 and 6 which are cone-tilting and Fermi-velocity corrections, respectively), of the corrections arising from the optical field only term 15 (the leading order contribution) has been reported in literature\cite{lin12}. The correspondence rule above generates several optical terms from the well known acoustic terms, $14$, and $17$ can be obtained from their acoustic counterparts $5$, and $6$ in this way. The two additional corrections (terms 17 and 18) have no acoustic counterpart, with the former representing a simple velocity correction proportional to the square of the optical deformation, with the latter an imaginary tensorial expression. The coupling of optical and acoustic field also generates Fermi velocity correction: term 24 in Table II, the hermitian pair of the imaginary opto-acoustic gauge term 23.

Finally, we consider corrections at second order in momentum, i.e. modification of the trigonal warping terms of pristine graphene. While contributions exist for all types (acoustic, optical, and opto-acoustic) we present in Table II the trigonal warping correction only for the optical field (terms 19 and 20). These terms are interesting as they depend directly on the optical field itself. We do not go beyond second order in momentum since, as will be discussed in the next section, the hermiticity of the effective Hamiltonian is guaranteed only up to second order in momentum. 

All of the expressions previously discussed have been obtained in the $\pi$-band approximation to the electronic structure of graphene. An interesting question is how these result change once both the $\sigma$ and $\pi$-hopping resulting form the bending of $p_z$ orbitals are included; this can be important for out-of-plane deformation\cite{Carbon}. While we do not present explicitly the analytical results we find the following: all terms that depend on the deformation tensor (either optical or acoustic) are universal, having identical form (but different numerical pre-factor) in either scheme. However, all the vector-valued terms generally change. In particular we find that that the compact vector expressions displayed in Table II no longer hold. The exception is the out-of-plane terms, which have exactly the same form in both schemes (but of course with some difference in the numerical pre-factor).

%%%%%%%%%%%%%%%%%%%%%%%%%%%%%%%%%%%%%%%%
% Hermiticity of effective Hamiltonian %
%%%%%%%%%%%%%%%%%%%%%%%%%%%%%%%%%%%%%%%%

\section{Hermiticity of the effective Hamiltonian}
\label{Sec-Herm}

The existence of imaginary gauge fields due to deformation in graphene was first discussed in Ref.~\onlinecite{PhysRevLett.108.227205} in which it was shown that the combination of the imaginary gauge from higher-order Cauchy-Born deformation (term 4 in Table II) and real Fermi velocity (term 6) acted to preserve hermiticity. As can be seen in Table II, the generalization to include both Cauchy-Born and non-Cauchy-Born deformation (and their coupling) generates a plethora of new terms in the effective Hamiltonian, both real and imaginary. In addition, imaginary gauge fields now also occur at leading order (for out-of-plane optical deformation). The question of the hermiticity of the effective Hamiltonian described by Table II must therefore be carefully examined.

It is useful to first consider the hermiticity of the following Hamiltonian:

\begin{equation}
\label{Gen_Ham}
H(\br,\bp)=\sum_{in} \frac{1}{n!} S_{in}(\br)\sigma_i\left(\frac{p}{\hbar}\right)^{n}
\end{equation}
where as before $n$ is the tuple $(n_1,n_2)$, $n!=n_1!n_2!$, $p^{n}=p_1^{n_1} p_2^{n_2}$. The fields $S_{in}$ are assumed complex valued and so this represents the most general Hamiltonian in which spatially varying fields couple to the momentum operator in $SU(2)$ space. In the specific case of graphene the $|n|=0$ terms represent scalar and gauge fields, the $|n|=1$ terms Fermi velocity correction, and the $|n|=2$ terms a correction to trigonal warping (in the tuple notation $|n|=n_1+n_2$).

It is obvious that in general Eq~\eqref{Gen_Ham} is not hermitian. 
In the following we will therefore first address the special conditions on the fields $S_{in}$ such that Eq~\eqref{Gen_Ham} is hermitian, before specializing to the case of deformed graphene where we show that, surprisingly, the special conditions that render Eq.~\eqref{Gen_Ham} hermitian are satisfied Cauchy-Born and non-Cauchy-Born deformations in graphene.

Our principle result is that Eq.~\eqref{Gen_Ham} is hermitian to all orders in momentum only if the spatial fields have no derivatives beyond second order.  For ``fast'' fields, beyond second order in spatial derivative, only terms up to second order in momentum are permitted. This represents a fundamental limitation on continuum theory: the transparency and efficiency over atomistic methods requires a price to be paid either in momentum by remaining close to an expansion point, or in deformation by having slowly varying fields.

\subsection{Hermiticity conditions from general Hamiltonian}\label{GH}

Term by term Eq.~(\ref{Gen_Ham}) violates hermiticity in potentially two ways: (i) the complex valued field terms and (ii) the coupling of spatial fields to the momentum operator. As we now show, these two can conspire together to restore hermiticity.

The requirement for hermiticity of Eq.~\eqref{Gen_Ham} can conveniently expressed in position representation as:

\begin{equation}
\label{phi1phi2}
 \int {\phi_1(\br)}^{\dagger} H(\br,\bp) {\phi_2(\br)} d{\br} 
 =\int (H(\br,\bp) \phi_1(\br))^{\dagger} {\phi_2(\br)} d{\br}.
\end{equation}
Let us first consider the right hand side of Eq.~\eqref{phi1phi2} for the real part of the fields with some arbitrary tuple $n$. The matrix element

\begin{equation}
\label{Real}
\frac{1}{n!}\langle\phi_1 |{\text Re}~ S_{in} \sigma_i p^{n} \phi_2\rangle
\end{equation}
can be integrated by parts, where application of the Leibniz rule then yields

\begin{equation}
\sum_{k \leq n}
\frac{(-1)^{|n|}}{n!}{{n}\choose{k}}\langle p^{n-k}\phi_1p^k {\text Re}~ S_{in}\sigma_i|\phi_2\rangle
\label{Her_Rel}
\end{equation}
(the condition of either a periodic or spatially localized deformation ensures the vanishing of the surface term).
The above equation can then be separated into a $k=0$ term, the hermitian counterpart of Eq.~(\ref{Real}),

\begin{equation}
\frac{(-1)^{|n|}}{n!}\langle p^{n}\phi_1 {\text Re} ~S_{in}\sigma_i|\phi_2\rangle,
\end{equation}
and the additional terms:

\begin{equation}
\frac{(-1)^{|n|}}{n!}\sum_{0<k\leq n}{{n}\choose{k}}\langle p^{n-k}\phi_1 p^k {\text Re}~ S_{in}\sigma_i|\phi_2\rangle.
\label{HR1}
\end{equation}
Thus Eq.~\eqref{Real}, by itself, violates hermiticity. However, let us now consider the imaginary part at one order lower in momentum. We choose a tuple $n'$ such that $|n'| = |n| - 1$ (with further condition to be specified subsequently):

\begin{equation}
\label{Imag}
\frac{i}{n'!}\langle\phi_1 |{\text {Im}}~ S_{in'} p^{n'}\sigma_i|\phi_2\rangle.
\end{equation}
Again, use of the Leibniz rule gives

\begin{equation}
\frac{(-1)^{|n|-1}i}{n'!}\sum_{k' \leq n'}{{n'}\choose{k'}}\langle p^{n'-k'}\phi_1 p^{k'} {\text {Im}}~ S_{in'}\sigma_i|
\phi_2\rangle.
\label{Her_Imag}
\end{equation}
The imaginary sign now generates two 
distinct types of hermiticity error as the hermitian counterpart of Eq.~\eqref{Imag}

\begin{equation}
-\frac{(-1)^{|n|-1}i}{n'!}\langle p^{n'}\phi_1 {\text {Im}}~ S_{in'}\sigma_i|\phi_2\rangle
\end{equation}
requires both correction at the same order of momentum $|n|-1$ due to the change in sign of the imaginary unit under conjugation:

\begin{equation}
\frac{2(-1)^{|n|-1}i}{n'!}\langle p^{n'}\phi_1 {\text{Im}}~ S_{in'}\sigma_i|\phi_2\rangle
\label{HI1}
\end{equation}
as well as the error from the Leibniz rule acting on the field term:

\begin{equation}
\frac{(-1)^{|n|-1}i}{n'!}\sum_{0<k'\leq n'}{{n'}\choose{k'}}\langle p^{n'-k'}\phi_1 p^{k'} {\text {Im}}~ S_{in'}\sigma_i|\phi_2\rangle.
\label{HI2}
\end{equation}
While two terms in Eqs.~(\ref{Real}) and (\ref{Imag}) are unsurprisingly not individually hermitian, in combination hermiticity can be restored. However, the two distinct types of errors generate generally incompatible conditions resulting in hermiticity restrictions on the fields of Eq.~\eqref{Gen_Ham}.

For order $O(p)$ we find from Eqs.~(\ref{HR1}) and (\ref{HI1}) that

\begin{equation}
\label{Her_Cond_1}
p^k ~{\text{Re}}~ S_{in}-2i~{\text{Im}}~S_{in'}=0
\end{equation}
with $|k|=1$ and $n'=n-k$. This in the case of graphene gives the hermiticity condition for the pair of real Fermi velocity and imaginary gauge field, and real cone tilting term and imaginary scalar field. A special case of this condition for Cauchy-Born deformation has been obtained in [\onlinecite{PhysRevLett.108.227205}].

For $O(p^{|k|})$ with $1<|k|<|n|-1$ we find from Eqs.~\eqref{HR1} and \eqref{HI2} the additional conditions

\begin{equation}
\label{Her_Cond_2}
\frac{1}{k!}p^k ~{\text{Re}}~ S_{in}- i\sum_{k' n'} \frac{1}{k'!}p^{k'}{\text{Im}}~S_{in'}=0
\end{equation}
where the sum is over all $n'$ and $k'$ that satisfy $n-k=n'-k'$, $|k'|=|k|-1$, and $|n'|=|n|-1$. We therefore require that the conditions in Eq.~(\ref{Her_Cond_2}) be obtainable from Eq.~(\ref{Her_Cond_1}) by differentiation. This is possible only for $|k| = 2$, i.e. only until the second order in momentum, beyond which the conditions become inconsistent and the Hamiltonian consequently non-hermitian.

The price to be paid for hermiticity must therefore be met either through momentum or through the field. For large momenta ($|n| > 2$) fields must be slowly varying; from Eq.~\eqref{Her_Cond_2} we see that derivatives higher than second order must vanish. Correspondingly, to have fast spatially varying fields momenta must be restricted to be second order or less. This, however, is simply what a general form of the Hamiltonian dictates. To determine if a given theory satisfies these conditions explicit forms of $S_{i n}(\br)$ must be supplied. For the special case of deformation in graphene, we now consider this question.

\subsection{Hermiticity conditions for deformation in graphene}\label{CH}

Using Eq.~(\ref{deformed}), we may write the field terms in Eq.~(\ref{Gen_Ham}) from each hopping function type $\eta$ as

\begin{equation}
S_{\eta i n}(\br)\sigma_i=\sum_{m} C_{\eta m}(\br)T_{\eta,m+n}.
\end{equation}
where $\eta$ labels the different types of hopping in sublattice space, with $\eta=1$ the Cauchy-Born obeying homogeneous deformation, and $\eta=2-4$ non-Cauchy-Born deformations, see Eq.~\ref{XXX}. As the associated sublattice space matrices were linearly independent, the hermiticity of each $\eta$ channel is independent from the others.

Substituting in Eq.~(\ref{Her_Cond_1}) and comparing the coefficients of different orders of $T_{\eta m}$, we find

\begin{eqnarray}
\label{C1}
\partial_x C_{\eta (|m|,0)}&=&2C_{\eta (|m|+1,0)},\nonumber\\
\partial_y C_{\eta (0,|m|)}&=&2 C_{\eta (0,|m|+1)}
\end{eqnarray}
and

\begin{equation}
\label{C2}
 \partial_x C_{\eta (|m|-l,l)}+\partial_y C_{\eta (|m|-l+1,l-1)}=2 C_{\eta (|m|+l-1,l)}
\end{equation}
for $l=1,|m|$, and where $|m| = m_1 +m_2$. Thus the hermiticity conditions for the effective Hamiltonian describing deformations in graphene depend only on the $C_{\eta m}$ coefficients, i.e. on the coefficients of the expansion of the change in the hopping function due to deformation. In Table \ref{t-table} it may be seen that all of the $C_{\eta m}$ indeed satisfy conditions Eq.~\eqref{C1} and Eq.~\eqref{C2} above; at higher orders ($|m|=3$ for acoustic and $|m|=2$ for optical) in deformation this is no longer possible.

Finally, for completeness we enumerate the various hermitian pairs in Table II. For the acoustic case, terms 3 and 4 are made hermitian by terms 6 and 5 respectively, while for the optical case, terms 10, 11, 12 have terms 15, 16 and 14 as their hermitian pairs. For opto-acoustic fields, the imaginary gauge (term 23) is made hermitian by Fermi velocity correction in term 25. Finally, the imaginary Fermi velocity terms arising from optical deformations (terms 17 and 18) are the  hermitian partners of the trigonal warping corrections to the field (terms 19 and 20 respectively).

%%%%%%%%%%%%%%%%%%%%%
% Numerical results %
%%%%%%%%%%%%%%%%%%%%%

\section{Numerical results}

\subsection{Details of the numerical method}
\label{NUM}

\begin{figure*}[!h]
\includegraphics[scale=0.9]{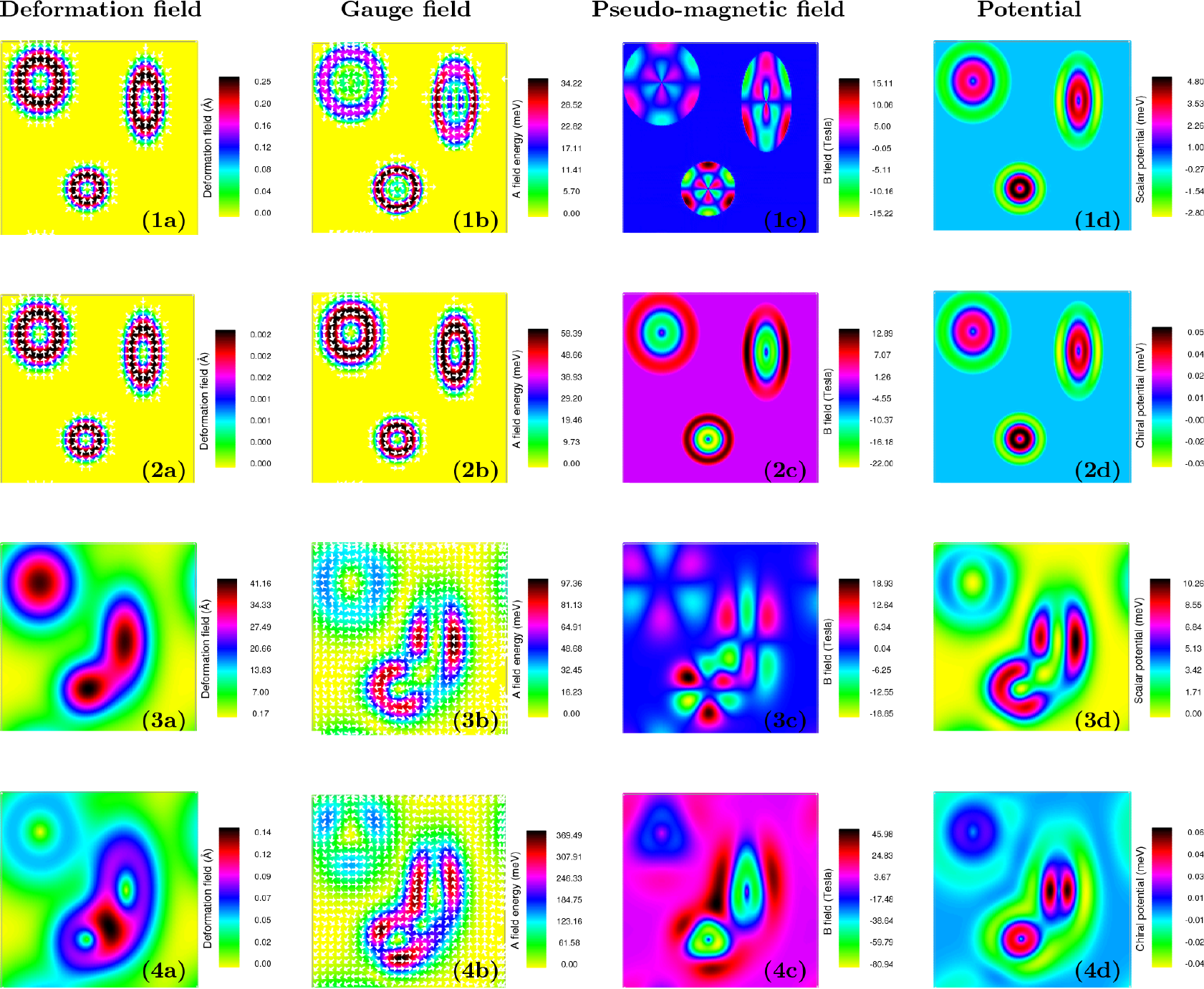}
\caption{\emph{Functional relationship between deformation and effective fields for Cauchy-Born and non-Cauchy-Born deformations}. Shown are (1a) in-plane acoustic, (2a) in-plane optical, (3a) out-of-plane acoustic, (4a) out-of-plane optical deformation fields with the corresponding effective fields the gauge field (second column), the pseudo-magnetic field (third column), and the scalar field  (fourth column). The scalar field is an electric potential for Cauchy-Born deformation and a chiral (mass generating) potential for non-Cauchy-Born deformation. Note that the effective fields in row 4 result from the combination of the deformation fields in (3a) and (4a), i.e., these effective fields result from a coupling of the out-of-plane acoustic and optical deformation fields. For the Cauchy-Born deformations (rows 1 and 3) the pseudo-magnetic field displays the clear influence of the underlying graphene lattice in the 6-fold petal structure of the field. In dramatic contrast the pseudo-magnetic field for the non-Cauchy-Born deformations follow closely the structure of the deformation field itself, and exhibit no symmetry lowering due to the underlying lattice. For further details on the nature of the deformations and the resulting effective fields, see Section~\ref{REL}. Calculations are performed at the K valley of the graphene BZ, and the sample area is $700a\times 700a$ with $a$ the lattice constant of graphene. At the K$^\ast$ valley the sign of the pseudo-gauge is reversed.}
\label{Fig2}
\end{figure*}

\begin{figure*}[!h]
\includegraphics[scale=0.9]{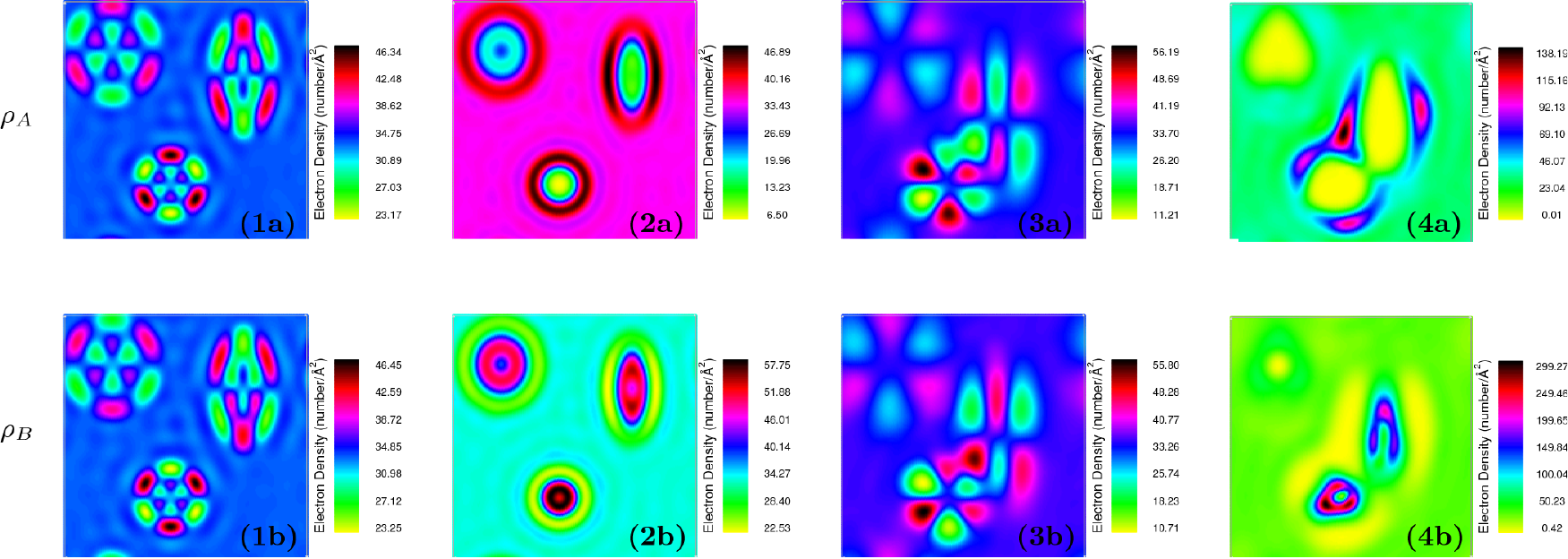}
\caption{Sublattice projected local density of states for the four types of deformations shown in Fig.~\ref{Fig2}. The acoustic (i.e., homogeneous Cauchy-Born) deformations generate  ``charge flowers'' with three of the ``petals'' exhibiting strong charge localization on the A sublattice with the other three petals strongly localized on the B sublattice. In dramatic contrast, the optical (i.e., non-Cauchy-Born) deformation generates ``charge walls'' and ``charge dots'' in which the A sub-lattice is localized at the perimeter of the deformation, and the B sub-lattice in the interior of the optical deformation field. In either case, the charge densities follow closely the pattern of the pseudo-magnetic fields induced by the deformation, see Fig.~\ref{Fig2}. Results are obtained by integrating an energy window from 0 to 100~meV; similar features are found in any energy window close to the Dirac point. Calculations are performed at the K valley of the graphene BZ.}
\label{Fig3}
\end{figure*}

\begin{figure*}[!h]
\includegraphics[scale=0.9]{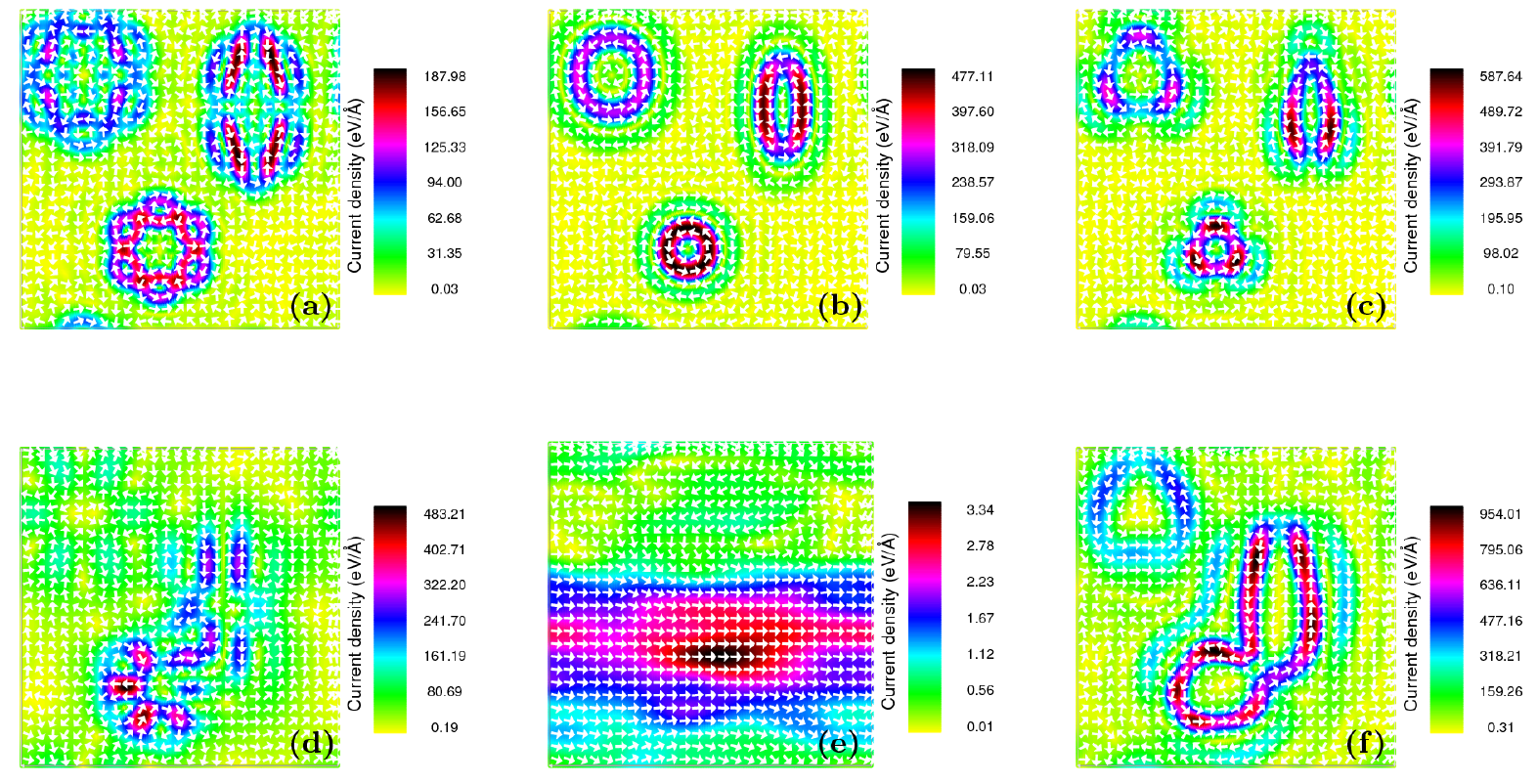}
\caption{Current densities due to homogeneous Cauchy-Born obeying deformations (acoustic) and non-Cauchy-Born deformations (optical), displaying the striking difference between current structure from these two types of deformations. Shown are current densities for all six possible cases of deformations: (a) acoustic in-plane, (b) optical in-plane, (c) opto-acoustic in-plane, (d) acoustic out-of-plane,  (e) optical out-of-plane and (f) opto-acoustic out-of-plane. Acoustic deformations give rise to local current loops\cite{currents2, currents3}, 3 clockwise and 3 anti-clockwise for a centro-symmetric deformation as seen in panel, but with more complex structure for the non-centro-symmetric cases, as seen in panels (a) and (d). In contrast, optical deformations, or deformations with an optical component, generate large scale current loops following the structure of the deformation, panels (b), (c), and (f). In all cases the underlying physical mechanism is snake states along nodal lines of the pseudo-magnetic field. The exception is the out-of-plane optical deformation, which is purely imaginary (see term 11 in Table~\ref{Table_bful}) and generates almost no current (reduced by two orders of magnitude as compared to all other cases), and exhibits almost no discernible connection to the deformation field. We attribute these small currents to the Fermi velocity anisotropy that results in a non-zero imaginary pseudo-magnetic field (see Section~\ref{REL}). Calculations are performed at the K valley of the graphene BZ, at the K$^\ast$ valley the sign of all currents is interchanged, as required by time reversal symmetry. The energy window of integration is between 0 and 100~meV, with similar results found in any energy window close to the Dirac point.
}
\label{Fig4}
\end{figure*}

\begin{table}
\begin{tabular}{|c|c|c|c|}
\hline
&& $t_{pp\sigma}(\bdel^2)$ & $t_{pp\pi}(\bdel^2)$\\
\hline
Full Slater-Koster scheme  & A & 63.4 & -28.59 \\
&B&1.5&1.5\\
\hline
H\"ukel model &A & - & -21.03\\
&B & - & 1.00 \\
\hline
\end{tabular}
\label{SKP}
\caption{Tight-binding schemes and parameters used in obtaining effective Hamiltonians. Shown are the parameterization of the $\sigma$- and $\pi$-hopping terms with an exponential $A e^{-B\bdel^2}$ form assumed ($\bdel$ is the hopping vector). The H\"ukel model restricts the hopping amplitude to only the second term of Eq.~\eqref{fullHF}, while the full Slater-Koster scheme retains both terms allowing for orbital-bending effects. }
\end{table}

For all the numerical results the hopping function has been chosen to be a Gaussian of the form

\begin{equation}
t(\bdel) = A \exp (-B \bdel^2).
\end{equation}
We choose this over the common alternative\cite{Gaussian, Electronic, Valley-filter2, Valley-filter4, Peeters-triaxial}

\begin{equation}
t(\bdel)=t_0 \exp[-\beta(|\bdel|/a_0-1)]
\end{equation}
as the latter is a slowly converging function in reciprocal space (it converges as $1/q^2$), due to the derivative discontinuity at the origin in real space.  This slow convergence makes the translation group sums in Eq.~\eqref{Heff2} slow to converge, and so we prefer the faster converging exponential form.

The full hopping function between the $p_z$ orbitals in the Slater-Koster scheme is given by

\begin{equation}
t(\delta^2) = \frac{\delta_z^2}{\delta^2}t_{pp\sigma}(\delta^2) + \biggl(1-\frac{\delta_z^2}{\delta^2}\biggr)t_{pp\pi}(\delta^2)
\label{fullHF}
\end{equation}
where the first term vanishes for planar materials such as pristine graphene, as well as for graphene with in-plane deformations. For out-of-plane deformations, the first term is non-zero, and for substantial out-of-plane deformation is important\cite{Carbon}. The values of the $A$ and $B$ parameters for both the $\pi$-hopping only H\"ukel model and for the full hopping function in Eq.~\eqref{fullHF} are given in Table~\ref{SKP}. The former is parameterized simply by requiring the nearest neighbour hopping to be $-2.8$~eV, and reproduction of the experimental Fermi velocity, for the full Slater-Koster tight-binding we use a parameterization developed in Ref.~\onlinecite{shall18}. In each case the translation group of the expansion point in Eq.~\ref{Heff2} is summed until the contribution of a star is $10^{-10}$ or less. Calculations are performed on a real space grid with derivatives obtained by FFT (the relative numerical error in the derivatives is found to be of the order of $10^{-8}$). For all calculations we numerically check the Hermiticity of the effective Hamiltonian, finding negligible errors in Hermiticity of $10^{-10} - 10^{-18}$ of the norm of the matrix. We typically use a $5\times5$ mesh of $\bk$-points for a square sample size of $700a\times700a$ with $a$ the lattice constant of graphene, and a corresponding $\bk$-vector density for other systems.

\subsection{Relation between deformation and effective fields}\label{REL}

Fundamental to the electronic theory of deformation is the functional relation between the deformation field and the pseudo-gauge and scalar potentials of the Dirac-Weyl equation. For acoustic (Cauchy-Born) deformation this relation is rather complex. For example, a centro-symmetric deformation $\bu^+(\br) = f(r) \hat{\bu}$ generates a pseudo-magnetic field $B_z^+ = g(r) \sin3\theta$. 
Evidently, the effective magnetic field depends not only on the applied deformation, but is entangled with the $C_3$ symmetry of the underlying lattice through the $\sin3\theta$ factor. This entanglement of lattice and deformation field occurs whatever the ratio of the length scale of the deformation to the lattice constant, and has profound consequences for the electronic structure of Cauchy-Born deformations. Two notable examples of this are (i) the necessity for special triaxial deformations\cite{Guinea2009}
to generate approximately uniform fields (essentially to ``undo'' the $C_3$ contribution of the lattice) and (ii) the occurrence of multiple local current loops within a deformed region of the lattice\cite{currents2, currents3}. This latter feature is very useful in the construction of valley filters\cite{Valley-filter, Valley-filter2, Valley-filter6, Valley-filter7, Valley-filter8}, but precludes the long range transport of valley polarized current by deformation as the current density will, in general, always consist of localized closed current loops incapable of transporting valley charge over extended distances.

Understanding the physics introduced by going beyond the Cauchy-Born rule therefore begins at the functional relationship between non-Cauchy-Born 
deformations and the resulting pseudo-gauge and scalar potential terms in the Hamiltonian. To this end we will consider a diverse set of deformation fields: (1) in-plane acoustic, (2) in-plane optical, (3) out-of-plane acoustic, and (4) a field with both out-of-plane optical and out-of-plane acoustic components. These are shown in panels (1a-4a) of Fig.~\ref{Fig2}. Of these only the pure acoustic deformations (1 and 3) obey the Cauchy-Born rule.

The deformation fields in panels (1a), (2a) have the form

\begin{equation}
\begin{cases}
\bu(\br) = \alpha \cos^2 \big(\frac{\pi |\beps(\br-\br_0)|}{2 R}\big) \hat{\br} & r < R,\\
0 & r > R
\end{cases}
\end{equation}
with $\alpha$, $\br_0$ and $R$ being the amplitude, center, and width respectively. $\beps$ is a matrix that can be used to transform the centro-symmetric deformation to one of lower symmetry. On the other hand, the out-of-plane deformation shown in panel (3a) is given by

\begin{equation}
u_z(r) = \beta \exp [-\{\beps(\br-\br_0)\}^2/a^2]
\label{GBUMP}
\end{equation}
with $\beta$, $\br_0$, and $a$ representing the amplitude, center and width of the deformation respectively. Finally the optical out-of-plane deformation field in panel (4a) is taken as the derivative of the deformation in (3a). In each of these panels can be seen three localized deformations, each with a different $\beps$ such that we consider not only centro-symmetric deformations but also lower symmetry cases. As can be seen, there is also some overlap between each of these three deformation fields with the total deformation field just given by the sum of all localized deformations.

We observe that the deformations in panels (1a) and (3a), which correspond to in-plane and out-of-plane acoustic fields respectively, generate the well known three-fold structure \cite{Electronic,PhysRevB.90.041411,multi-uni} of the pseudo-magnetic field, panels (1c) and (3c). This is found in both the high and low symmetry local deformation fields. On the other hand, the in-plane optical and and out-of-plane opto-acoustic deformations, panels (2a) and (4a) respectively, generate very different pseudo-magnetic fields. Here the pseudo-magnetic field does not exhibit a three-fold structure, but instead follows closely the deformation field. This represents a fundamental difference between Cauchy-Born and non-Cauchy-Born deformations in graphene. Several consequences that follow from this distinction will be presented subsequently, however it immediately implies that a triaxial deformation will not be required to create an approximately uniform magnetic field for non-Cauchy-Born deformations. 

These qualitative differences do not manifest themselves in the scalar fields, however, which in each case follows the deformation field as can be seen in panels (1d-4d) of Fig.~\ref{Fig2}. These fields, however, play a much less important role than the pseudo-magnetic fields.

To better understand these features we can consider the leading order contribution to 
the pseudo-magnetic and scalar fields plotted in Fig.~\ref{Fig2}. The pseudo-
magnetic field generated by an in-plane optical deformation, $B_s^-$, and that 
generated by a the coupling of out-of-plane optical and acoustic fields, $B_s^{\pm}$, can for arbitrary deformation be written as 

\begin{eqnarray}
B_s^- & = & \bnab . \bu^-, \label{BDIV} \\
B_s^{\pm} & = & \bnab u_z^-.\bnab u_z^+ + u_z^- \bnab^2u_z^+ \label{BCUR}
\end{eqnarray}
(these are terms 8 and 22 in Table II
respectively). Thus these pseudo-magnetic fields depend only on the deformation fields $\bu^-(\br)$ and $u^\pm_z$, while a corresponding formula cannot be written down for acoustic deformations, explaining the quite different functional relation between deformation field and pseudo-gauge for these two cases. Similar formulae may be written down for the the scalar fields.

Finally, the out-of-plane optical field is very different from all of the above. This follows from the fact that the leading order gauge in this case is pure imaginary. Interestingly, this imaginary gauge also produces an imaginary magnetic field. Such imaginary gauges have been discussed before in the context of the imaginary acoustic gauge (term 4 in Table II), but not the curious fact of an imaginary magnetic field in the context of a hermitian Hamiltonian. To understand the presence of an imaginary magnetic field we recall that any imaginary gauge $\bGam$ is paired with a real Fermi velocity correction $v_F^{ij}$ (see Section \ref{Sec-Herm}) according to Eq.~(\ref{Her_Cond_1}), which can conveniently be rewritten as

\begin{equation}
\hbar\partial_j v_F^{ij}(\br)\sigma_i + 2 \Gamma_i\sigma_i = 0.
\end{equation}
If the tensor $v_F^{ij}$ is isotropic, i.e. $v_F^{ij} = v_F \delta_{ij}$ then this equation may be recast as

\begin{equation}
{\bf{\Gamma}} = -\frac{1}{2}\hbar \bnab v_F(r),
\end{equation}
implying no imaginary pseudo-magnetic field as gauge is then irrotational. In this case, as has been discussed in Ref.~[\onlinecite{PhysRevB.87.165131}], the imaginary gauge field corresponds to local pseudospin rotations. However, for non-isotropic Fermi velocities, which is usually the case, there will be a non-vanishing imaginary pseudo-magnetic field.

\subsubsection{Electronic consequences}

We first consider the local electron density projected onto the A and B sub-lattice generated by each of the deformations shown in Fig.~\ref{Fig2}. As may be seen in Fig.~\ref{Fig3}, a striking difference between Cauchy-Born and non-Cauchy-Born deformations is found. The acoustic deformations generate ``charge flowers'' in which 3 petals are A sublattice localized, and 3 B sublattice localized, seen for both the high and low symmetry deformations. The optical and opto-acoustic deformations, however, generate a very different pattern of charge localization, with A sublattice localization on the perimeter of the deformation, and B sublattice localization on the interior of the deformation. In each case, as may be seen via comparison with Fig.~\ref{Fig2}, the localization follows closely the pseudo-magnetic field $B_s$, with A (B) localization on positive (negative) regions of $B_s$. These distinct patterns of sub-lattice polarization could be used to probe the presence of non-Cauchy-Born deformation in experiment. Indeed, such charge separation between the sublattices, which can be viewed as pseudo-spin polarization due to a pseudo-Zeeman field, represents a useful local probe of deformation
Interestingly, we also see from Fig.~\ref{Fig3} that while the pseudospin polarization integrates to zero over the sample for acoustic deformations, this is not true for optical deformations which, within the energy window of integration, displays a net pseudospin moment. Note however that the sign of the pseudo-spin moment will be opposite at the K and K$^\ast$ valleys.

Current carrying snake states at nodal lines of the pseudo-magnetic field are central to straintronics in graphene \cite{currents2, currents3, Electronic, Nikodem, Valley-filter, Valley-filter2, Valley-filter6, Valley-filter7, Valley-filter8},
and so we now examine the nature of the current carrying states arising from Cauchy-Born and non-Cauchy-Born deformations.
In Fig.~\ref{Fig4} we plot the current densities for all possible types of deformations. For acoustic in-plane and out-of-plane deformations (panels (a) and (d)) one finds local current loops\cite{eddy_currents,currents2,currents3} confined at the nodes of the pseudo-magnetic field along zigzag directions, thereby causing the current density to loop around the petals of flower structure shown in Fig.~\ref{Fig2}(1c) and (3c). On the A sub-lattice, the current flows clockwise whereas it is anticlockwise for the B sub-lattice\cite{currents2,currents3}. The origin of these current density patterns are snake states\cite{1998,1999,2000, 2008, 2011, 2015}, generated by the reversal of cyclotron motion as a charge carrier crosses a node of the pseudo-magnetic field. For non-Cauchy-Born deformations the nodal structure of the pseudo-magnetic field is dramatically different, given simply by the zeros of the divergence of the optical deformation field (see Eq.~\eqref{BDIV}), or by zeros in the curvature for out-of-plane opto-acoustic deformation (see Eq.~\eqref{BCUR}). Nodal lines will therefore follow the basic geometry of the deformation field, and the corresponding snake states generate local currents flowing along topographic features of the deformation. This can clearly be seen in  panels (b), (c) and (f) corresponding to in-plane optical, in-plane opto-acoustic and out-of-plane opto-acoustic, respectively, where the local current simply loops around the boundary of the deformation. This qualitative difference in the form of the local current densities has profound implications for ``straintronics'' applications. While the loop structure of acoustic deformation fields can be used to design valley filters, the extended nature of the snake states for optical and opto-acoustic deformation allows, in contrast, the possibility of valley polarized charge transport.
Finally, we note that the currents in panel (e) due to out-of-plane optical deformation, evidently very different to all the other cases, are negligible due to leading order of the gauge field (term 11 in Table~\ref{Table_bful}) being imaginary.

\begin{figure}[!h]
\includegraphics[width=0.4\textwidth]{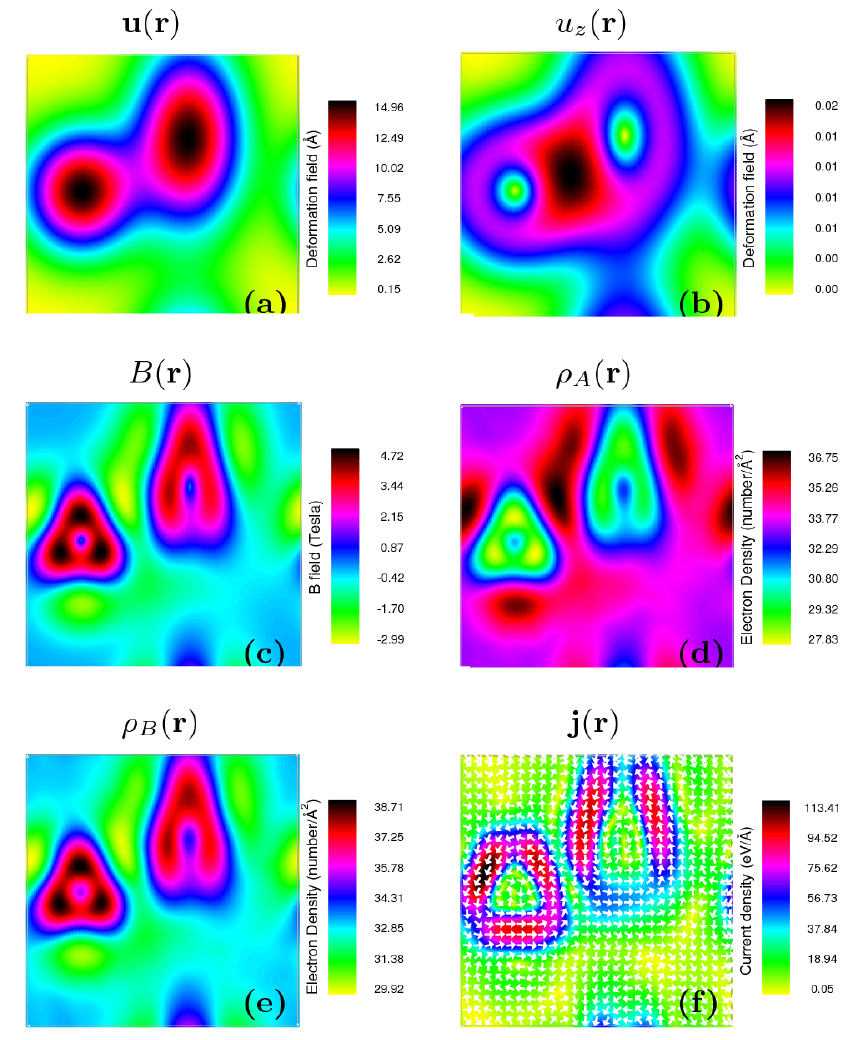}
\caption{Impact of inclusion of $\sigma$-hopping on the coupling between in- and out-of-plane deformations. Panels (a) and (b) display the in-plane and out-of-plane deformation fields, and panel (c) the resulting pseudo-magnetic field. In panels (d-f) are shown, respectively, the sub-lattice projected density and the current, integrated in an energy window from 0 to 100~meV at the K valley (similar results are found at other energy windows close to the Dirac point). As can be seen by comparison with Figs.~\ref{Fig2}-\ref{Fig4}, which include only $\pi$-hopping, the qualitative physics of the pseudo-gauge following the deformation is unchanged. The sample size is $700a\times700a$ with $a$ the lattice constant of graphene.}
\label{NF} 
\end{figure}

An important point to consider is how these results change upon inclusion of $\sigma$-hopping. This can be of relevance to out-of-plane deformation, the so called orbital-bending effect\cite{Carbon}, and to examine this point in Fig.~\ref{NF} we present calculations for out-of-plane deformation using effective gauge and scalar fields derived using the full Slater-Koster scheme (see Eq.~\eqref{fullHF} in Section \ref{NUM}) rather than the $\pi$-hopping only H\"ukel model. As has been noted, employing the full Slater-Koster scheme changes only the coefficients of the order $O(\epsilon)$ terms in Table \ref{TAB}, while preserving the forms, but does generate a plethora of new second order terms between out-of-plane and in-plane deformation. Thus the most ``vulnerable'' result of the preceding discussion is the gauge fields resulting from the coupling of in- and out-of-plane deformations. However, as can be seen from Fig.~\ref{NF}, the qualitative physics is reassuringly unchanged.

\subsection{A non-linear Gaussian bump}

\begin{figure}[!h]
\includegraphics[scale=0.9]{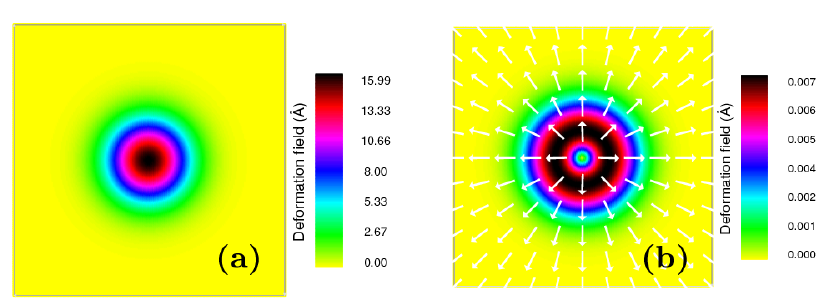}
\caption{The deformation field employed in the non-linear Gaussian bump of Fig.~\ref{Fig5}, which consists of (a) an out-of-plane acoustic deformation as a Gaussian bump of height around 40 \si{\angstrom} and (b) an in-plane optical deformation in the form of a Gaussian ring (obtained as the scaled derivative of the acoustic field). The sample size is $700a\times700a$ with $a$ the lattice constant of graphene.}
\label{DF}
\end{figure}

\begin{figure*}[!h]
\includegraphics[scale=0.9]{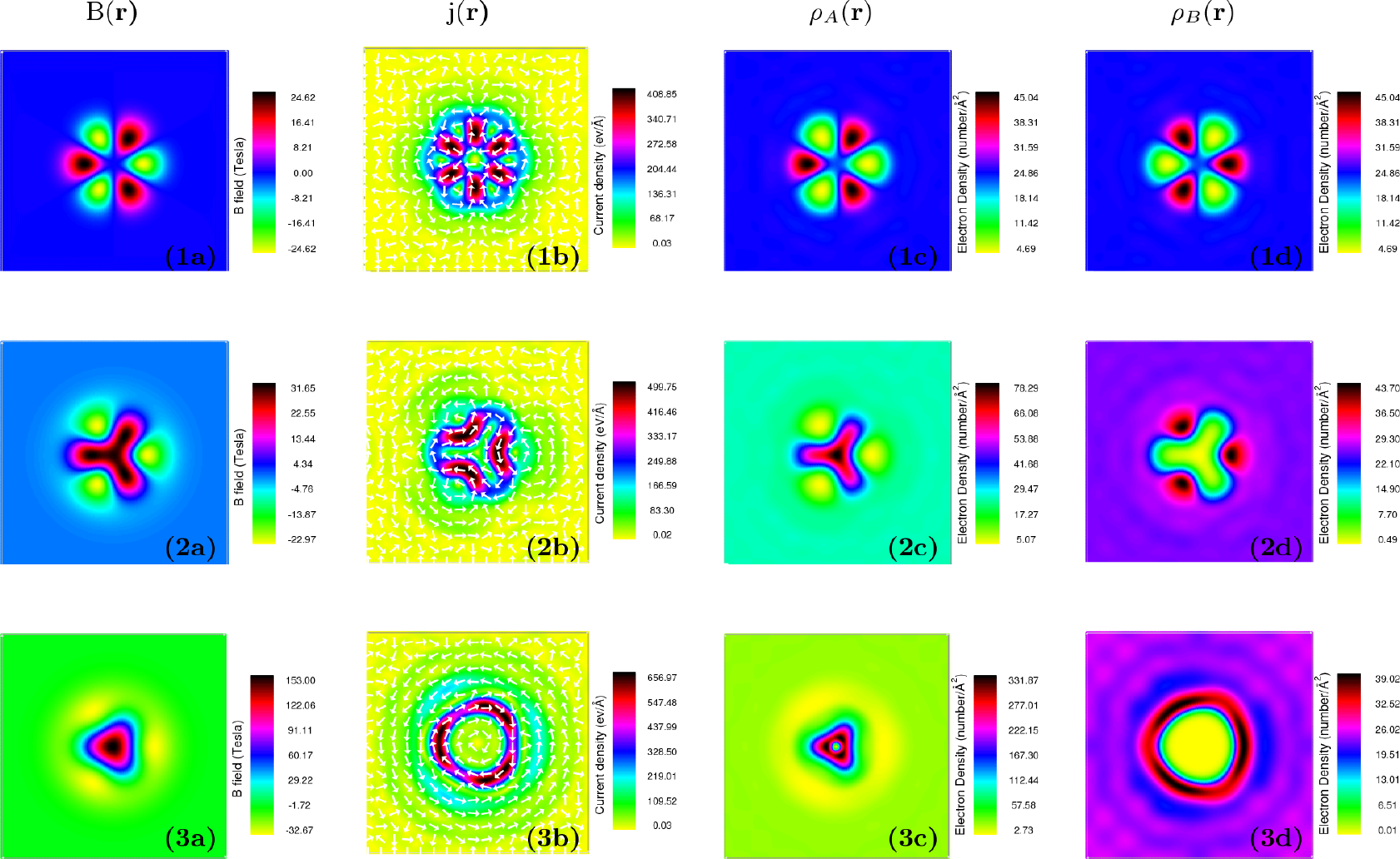}
\caption{\emph{The non-linear Gaussian bump}. Shown are the (a) pseudo-magnetic fields, (b) current densities, and projected charge densities on the A sublattice (c) and the B sublattice (d) for a Gaussian bump with only acoustic Cauchy-Born deformation (first row), but in the second and third rows with the addition on an in-plane optical deformation of maximum magnitude $10^{-3}$~\si{\angstrom} (second row) and $7\times 10^{-3}$~\si{\angstrom} (third row). The Cauchy-Born deformation field is shown in Fig.~\ref{DF}(a), with the optical non-Cauchy-Born component for the last row in Fig.~\ref{DF}(b). Evidently, the inclusion of an optical component in the deformation field has a dramatic effect on all physical quantities, with notably the 6-fold local current pattern of the Cauchy-Born deformation replaced by a single current density circulating around the deformation, and a striking change in the pseudospin polarization from the ``charge flowers'' seen in panels (1c) and (1d) to a polarization between the perimeter and interior of the deformed region seen in panels (3c) and (3d). The energy window of integration is 0 to 100~meV, but similar results are found in any energy window close to the Dirac point. The sample size is $700a\times700a$ with $a$ the lattice constant of graphene.}
\label{Fig5}
\end{figure*}

The Drosophila melanogaster of the theory of deformation in graphene is the Gaussian bump\cite{Gaussian, currents2, Electronic, multi-uni}. Here we wish to examine how the electronic structure this prototypical localized deformation is modified if the deformation becomes fast on the scale of the lattice constant; a non-linear Gaussian bump. To this end we will add to the acoustic field of the Gaussian bump a successively stronger in-plane optical field, given by a scaled derivative of the acoustic field.
In Fig.~\ref{DF} we display the form of both these deformation fields. Note that the scalar product of the deformation fields in the opto-acoustic coupling (term 22 in Table II) ensures that in this example the optical and acoustic fields do not couple in the electronic structure.

As may be seen in Fig.~\ref{Fig5}, with only a small optical component, the changes in the pseudo-magnetic field, current densities and charge densities are dramatic. We see that the ``flower structure'' of the pseudo-magnetic field changes first by the joining together of the positive $B_s$ petals at the centre of the deformation, with a corresponding repulsion of the negative $B_s$ petals, panel (2a), and then to a structure in which $B_s$ simply changes sign between the interior and perimeter of the deformation, panel (3a). The local current loops and sublattice polarization correspondingly change, with the acoustic ``charge flowers'' and associated 6-fold current loops replaced by the interior/perimeter sublattice polarization noted in the previous section, and the current density simply circulating around the perimeter of the non-linear Gaussian bump.

\subsection{Optical quenching in one dimensional deformations}

\begin{figure}
\includegraphics[scale=0.3]{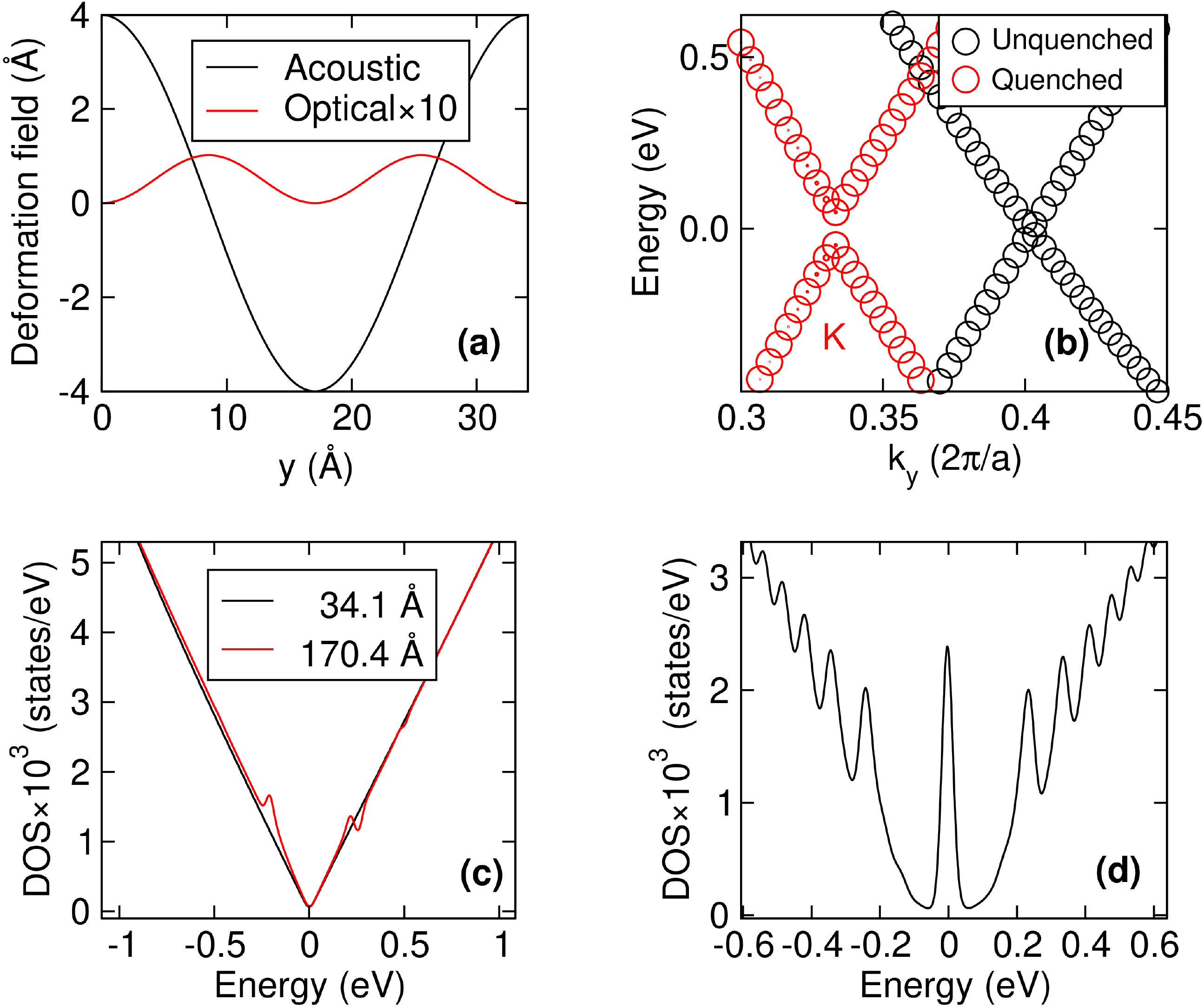} 
\caption{\emph{Optical quenching of armchair corrugations in graphene}: In panel (a) the acoustic and optical deformation fields are shown, see Eq.~\ref{acou} and Eq.~\ref{opt} respectively, representing a corrugation of the graphene lattice in the armchair direction. The acoustic deformation is taken from Ref.~[\onlinecite{midgap}] (b) The acoustic field generates a gauge field that displaces the Dirac cone off the high symmetry K point, dark (black) symbols, while addition of the optical deformation field exactly compensates this gauge resulting in the return of the Dirac cone to the K point, light (red) symbols, thus quenching the effect of the acoustic deformation. (c) The density of states (DOS) for a corrugation period of $170.4$~\si{\angstrom}, showing that optical quenching in addition can remove Landau levels (LL) that would be seen in the absence of the optical deformation field. (d) Optical deformation can by itself create Landau levels; shown is a $700 a$ corrugation with $a$ the lattice constant of graphene (in the zigzag direction) resulting in a clear sequence of chiral Landau levels ($E_n\propto\sqrt{n}$, $n\in\mathbb{N}$).}
\label{2}
\end{figure}

A prototype of the flexural ripples that occur in graphene is the one dimensional out-of-plane corrugation\cite{midgap, midgap2, 77K-ridge, PhysRevB.96.241405, PhysRevB.87.205405, 0295-5075-84-1-17003, Quantized}, typically taken to be a sinusoidal form. If the pseudo-magnetic field ($B_s$) changes slowly on the scale of the cyclotron length, such deformations result in a spectrum of Landau levels, whereas in the opposite limit the Dirac cone is displaced from the high symmetry K point\cite{Carbon}. The amplitude of $B_s$ is also very sensitive to the direction of the corrugation - in the zigzag directions it vanishes, with the maximum amplitude found along the armchair directions. Thus an imposed sinusoidal corrugation in graphene generates no effect in the zigzag direction, but a pronounced effect when in the armchair direction. Remarkably, \emph{ab-initio} calculations\cite{midgap,Carbon} reveal that upon allowing atomic relaxation to occur the situation is reversed: a corrugation in the armchair direction neither shifts the Dirac cone\cite{midgap} or generates LLs\cite{midgap,Carbon}, while in the zigzag direction LL type structures are clearly seen in the density of states\cite{Carbon}.

The pseudo-gauge in such corrugation deformations is often understood solely in terms of the contribution arising from acoustic (i.e. homogeneous Cauchy-Born) deformation. However, both acoustic and optical deformations give rise to gauge terms. For a simple one dimensional deformation one might suppose atomic relaxation to create an optical $B_s^-$ exactly canceling the imposed acoustic $B_s^+$: this would result in a large reduction in the system energy. This could occur in the armchair direction, and would explain the vanishing of the effects of deformation upon atomic relaxation, but could not occur in the zigzag direction as there $B_s^+$ vanishes anyway. The remaining non-zero $B_s^-$ could then generate the observed LL type structure seen in the density of states. In this way the presence of optical deformations may explain the unexpected switching between the natures of the armchair and zigzag directions upon atomic relaxation. To see if this is plausible we first consider an out-of-plane sinusoidal acoustic deformation given by

\begin{equation}
u_z^+= u_0^+ \cos (k_y^+ y),
\label{acou}
\end{equation}
with amplitude $u_0^+$ and wavenumber $k_y^+$, the corresponding gauge (term 2 in the Table \ref{Table_bful}) is

\begin{equation}
\label{AA}
A_x^+ = -\frac{\beta^+}{2} (u_0^+)^2 (k_y^+)^2 {\sin}^2 (2 k_y^+ y),
\end{equation}
with $\beta^+$ being the coefficient associated with the gauge fields from acoustic deformation. Now, we assume that the relaxation is given by a sinusoidal optical deformation

\begin{equation}
 u_y^-= u_0^- \cos (k_y^- y),
 \label{opt}
\end{equation} 
with unknown amplitude $u_0^-$ and wavenumber $k_y^-$. The gauge field corresponding to this optical deformation is given by (term 8 in table \ref{Table_bful})

\begin{equation}
 A_x^-=\beta^-u_0^-\cos (k_y^- y).
\end{equation}
By requiring that this gauge field exactly compensate the acoustic gauge field, Eq.~\eqref{AA}, we may solve for the unknown $u_0^-$ and $k_y^-$. In this way we find

\begin{eqnarray}
u_0^- & = &\frac{1}{2}\frac{\beta^+}{\beta^-}(u_0^+)^2 (k_y^+)^2 \label{A-quench} \\
k_y^- & = & 2k_y^+. \label{A-quench1}
\end{eqnarray}

To access the plausibility of this ``optical quenching'' mechanism, we consider the results of Ref.~[\onlinecite{midgap}]. In this work a sinusoidal deformation of $\lambda_y^+ = 34$~\si{\angstrom} and amplitude $u^+ = 4$~\si{\angstrom}, created a Dirac point shift of $0.042\frac{2\pi}{a}$. By applying an identical deformation (see Fig.~\ref{2}a) we find $0.069\frac{2\pi}{a}$ (see Fig.~\ref{2}b), a reasonable agreement given that we neglect both the $sp^2$ bands and $t_{pp\sigma}$ contribution to the hopping within the $\pi$-band. For these parameters of the acoustic deformation, Eqs.~\eqref{A-quench} and \eqref{A-quench1} yield an optical deformation of amplitude 0.1~\si{\angstrom} and wavelength exactly half that of the imposed optical deformation. Inspection of Fig.~1 in Ref.~[\onlinecite{midgap}], reveals that this describes very well the impact of atomic relaxation: the period of the optical deformation exactly matches, with the amplitude in reasonable agreement.

Optical quenching also offers a plausible explanation for the absence of LLs in the armchair direction, as increasing the wavelength of the corrugation results, by Eq.~\eqref{A-quench}, in a reduction of the amplitude of the quenching optical field. Thus only small relaxation effects are required to quench the pseudo-magnetic fields generated by the long wavelength corrugations that give rise to Landau levels. This is illustrated in Fig.~\ref{2}c where, for a long wavelength case that does generate Landau levels in the absence of lattice relaxation\cite{midgap}, the density of states is found to be free of Landau levels upon inclusion of optical deformation. 

In contrast to the suppression of the effects of deformation for armchair corrugations, in \emph{ab-initio} calculations quite significant changes in the density of states are seen for corrugation in the zigzag direction\cite{Carbon}. As was mentioned above, this result is also consistent with the optical quenching scenario described here, as in this direction the acoustic gauge is identically zero while the optical gauge (as can be seen term 11 of Table II) is independent of the direction of the applied corrugation. To illustrate this in Fig.~\ref{2}(d) we demonstrate the creation of LL via a pure optical corrugation in the zigzag direction.

\subsection{Corrections to effective fields derived using the Cauchy-Born rule}

Finally we wish to make contact between the work here and two studies that have investigated the consequences of going beyond the Cauchy-Born rule via an energy minimization procedure that fixes the relation between the optical and acoustic deformation fields\cite{non-uniform,non-uniform2}. Given this relation between these two fields, it was then shown that the pseudo-gauge is renormalized\cite{non-uniform} as compared to the standard results obtained within the Cauchy-Born rule, while the form of Fermi velocity is changed\cite{non-uniform2}. The purpose of this section is simply to demonstrate that, given the same relation between these two fields as an input, the results of Table II are in complete concordance with the results of these studies.

The optical deformation field resulting from a given acoustic field $\bu^+(\br)$ was found to be\cite{non-uniform}

\begin{equation}
\bu^- = \alpha(2 \epsilon_{xy}^+, \epsilon_{xx}^+-\epsilon_{yy}^+)
\end{equation}
Using this relation in conjunction with the lowest order contributions to the gauge and Fermi velocity due to optical deformation (terms 8 and 15 in Table II) we find for the pseudo-gauge field

\begin{equation}
\beta(\epsilon_{xx}^+-\epsilon_{yy}^+, -2 \epsilon_{xy}^+) 
\label{OA}
\end{equation}
and for Fermi velocity correction

\begin{equation}
\gamma \begin{pmatrix}
\epsilon_{xx}^+-\epsilon_{yy}^+ & 2 \epsilon_{xy}^+ \\
2 \epsilon_{xy}^+ & -(\epsilon_{xx}^+-\epsilon_{yy}^+)
\end{pmatrix}
\end{equation} 
with $\alpha, \beta, \gamma$ being constants. These have precisely the same form as obtained in Refs.~[\onlinecite{non-uniform}] and [\onlinecite{non-uniform2}] respectively.

Interestingly, for the ``optical quenching'' described in the previous section (and in the ab-initio results of Ref.~[\onlinecite{midgap}]), the period of the optical deformation resulting from atomic relaxation is exactly half that of the imposed acoustic deformation, showing that under heavy loading of graphene the relation between atomic displacement and deformation field described in Eq.~\eqref{OA} breaks down.

\section{Conclusions and discussion}

The Cauchy-Born (CB) rule states that around any material point deformation is homogeneous, and   understanding how deformation modifies the electronic properties of graphene has largely been based on this assumption. However, for non-Bravais crystals the CB rule is known to  break down, and \emph{ab-initio} simulations indicate that this is indeed the case in graphene. We have therefore generalized the continuum (Dirac-Weyl based) theory of deformation in graphene to include both Cauchy-Born deformation, described by a continuous acoustic field $\bu^+(\br)$, and deformations involving the microscopic degrees of freedom associated with the two sublattices, encoded in an optical displacement field $\bu^-(\br)$.

Employing an exact mapping of the Slater-Koster tight-binding method onto a continuum Hamiltonian $H(\br,\bp)$ we are able to treat both these fields on an equal footing and find the optical field, and the coupling of optical and acoustic fields, introduces qualitatively new pseudo-gauge and chiral fields to the Dirac-Weyl equation. Our theory, as it must, also reproduces all the well known pseudo-gauge, scalar potential, Fermi velocity, and cone-tilting corrections that the homogeneous continuum theory finds.

At the heart of the physics of lattice distortion in graphene is the functional relation between deformation and the effective Dirac-Weyl fields they generate, and we have shown that this is profoundly different for homogeneous Cauchy-Born deformations as compared to non-Cauchy-Born deformations. In the former, as is well known, the pseudo-magnetic field is ``entangled'' with the underling lattice of graphene\cite{Gaussian, multi-uni, Electronic, Local, Valley-filter2, currents2, currents3}. However for non-Cauchy-Born deformations the gauge field depends only on the deformation field. There are two consequences of this fact that appear striking. While homogeneous deformations result in a current density $\bj(\br)$ of multiple closed loops\cite{currents2, currents3}, in non-Cauchy-Born deformations $\bj(\br)$ follows topographic features of the deformation, e.g. nodal lines in the curvature. While the former feature can be utilized to design valley filters through local deformation ``bumps''\cite{Valley-filter2}
; the latter feature in principle allows for the transport of valley polarized charge over arbitrarily long distances e.g. along a designed ridge - a complementary and useful feature for straintronics. Secondly, to create an approximately uniform magnetic field from homogeneous deformation requires ``disentangling the lattice'' via a compensating triaxial deformation\cite{Guinea2009}, unnecessary for non-Cauchy-Born deformations. While the triaxial deformation field appears as a natural deformation in suitably sized nanobubbles\cite{Science-NB}, atomistic simulation suggests an important role for lattice relaxation\cite{Peeters-triaxial, Carbon, midgap}. Deformations beyond the Cauchy-Born rule may thus play a complementary role in creating a uniform magnetic field.

%\bibliography{deformation}

\begin{thebibliography}{70}
\expandafter\ifx\csname natexlab\endcsname\relax\def\natexlab#1{#1}\fi
\expandafter\ifx\csname bibnamefont\endcsname\relax
  \def\bibnamefont#1{#1}\fi
\expandafter\ifx\csname bibfnamefont\endcsname\relax
  \def\bibfnamefont#1{#1}\fi
\expandafter\ifx\csname citenamefont\endcsname\relax
  \def\citenamefont#1{#1}\fi
\expandafter\ifx\csname url\endcsname\relax
  \def\url#1{\texttt{#1}}\fi
\expandafter\ifx\csname urlprefix\endcsname\relax\def\urlprefix{URL }\fi
\providecommand{\bibinfo}[2]{#2}
\providecommand{\eprint}[2][]{\url{#2}}

\bibitem[{\citenamefont{Roldán et~al.}(2015)\citenamefont{Roldán,
  Castellanos-Gomez, Cappelluti, and Guinea}}]{rol15}
\bibinfo{author}{\bibfnamefont{R.}~\bibnamefont{Roldán}},
  \bibinfo{author}{\bibfnamefont{A.}~\bibnamefont{Castellanos-Gomez}},
  \bibinfo{author}{\bibfnamefont{E.}~\bibnamefont{Cappelluti}},
  \bibnamefont{and} \bibinfo{author}{\bibfnamefont{F.}~\bibnamefont{Guinea}},
  \bibinfo{journal}{Journal of Physics: Condensed Matter}
  \textbf{\bibinfo{volume}{27}}, \bibinfo{pages}{313201}
  (\bibinfo{year}{2015}),
  \urlprefix\url{http://stacks.iop.org/0953-8984/27/i=31/a=313201}.

\bibitem[{\citenamefont{Amorim et~al.}(2016)\citenamefont{Amorim, Cortijo,
  de~Juan, Grushin, Guinea, Gutiérrez-Rubio, Ochoa, Parente, Roldán, San-Jose
  et~al.}}]{AMORIM20161}
\bibinfo{author}{\bibfnamefont{B.}~\bibnamefont{Amorim}},
  \bibinfo{author}{\bibfnamefont{A.}~\bibnamefont{Cortijo}},
  \bibinfo{author}{\bibfnamefont{F.}~\bibnamefont{de~Juan}},
  \bibinfo{author}{\bibfnamefont{A.}~\bibnamefont{Grushin}},
  \bibinfo{author}{\bibfnamefont{F.}~\bibnamefont{Guinea}},
  \bibinfo{author}{\bibfnamefont{A.}~\bibnamefont{Gutiérrez-Rubio}},
  \bibinfo{author}{\bibfnamefont{H.}~\bibnamefont{Ochoa}},
  \bibinfo{author}{\bibfnamefont{V.}~\bibnamefont{Parente}},
  \bibinfo{author}{\bibfnamefont{R.}~\bibnamefont{Roldán}},
  \bibinfo{author}{\bibfnamefont{P.}~\bibnamefont{San-Jose}},
  \bibnamefont{et~al.}, \bibinfo{journal}{Physics Reports}
  \textbf{\bibinfo{volume}{617}}, \bibinfo{pages}{1 } (\bibinfo{year}{2016}),
  ISSN \bibinfo{issn}{0370-1573}, \bibinfo{note}{novel effects of strains in
  graphene and other two dimensional materials},
  \urlprefix\url{http://www.sciencedirect.com/science/article/pii/S0370157315005402}.

\bibitem[{\citenamefont{Shallcross et~al.}(2017)\citenamefont{Shallcross,
  Sharma, and Weber}}]{shall17}
\bibinfo{author}{\bibfnamefont{S.}~\bibnamefont{Shallcross}},
  \bibinfo{author}{\bibfnamefont{S.}~\bibnamefont{Sharma}}, \bibnamefont{and}
  \bibinfo{author}{\bibfnamefont{B.~H.} \bibnamefont{Weber}},
  \bibinfo{journal}{Nature Communications} \textbf{\bibinfo{volume}{8}},
  \bibinfo{pages}{342} (\bibinfo{year}{2017}), ISSN \bibinfo{issn}{2041-1723}.

\bibitem[{\citenamefont{Suzuura and Ando}(2002)}]{2002-paper}
\bibinfo{author}{\bibfnamefont{H.}~\bibnamefont{Suzuura}} \bibnamefont{and}
  \bibinfo{author}{\bibfnamefont{T.}~\bibnamefont{Ando}},
  \bibinfo{journal}{Phys. Rev. B} \textbf{\bibinfo{volume}{65}},
  \bibinfo{pages}{235412} (\bibinfo{year}{2002}),
  \urlprefix\url{https://link.aps.org/doi/10.1103/PhysRevB.65.235412}.

\bibitem[{\citenamefont{Ma\~nes}(2007{\natexlab{a}})}]{ref13}
\bibinfo{author}{\bibfnamefont{J.~L.} \bibnamefont{Ma\~nes}},
  \bibinfo{journal}{Phys. Rev. B} \textbf{\bibinfo{volume}{76}},
  \bibinfo{pages}{045430} (\bibinfo{year}{2007}{\natexlab{a}}),
  \urlprefix\url{https://link.aps.org/doi/10.1103/PhysRevB.76.045430}.

\bibitem[{\citenamefont{de~Juan et~al.}(2012)\citenamefont{de~Juan, Sturla, and
  Vozmediano}}]{PhysRevLett.108.227205}
\bibinfo{author}{\bibfnamefont{F.}~\bibnamefont{de~Juan}},
  \bibinfo{author}{\bibfnamefont{M.}~\bibnamefont{Sturla}}, \bibnamefont{and}
  \bibinfo{author}{\bibfnamefont{M.~A.~H.} \bibnamefont{Vozmediano}},
  \bibinfo{journal}{Phys. Rev. Lett.} \textbf{\bibinfo{volume}{108}},
  \bibinfo{pages}{227205} (\bibinfo{year}{2012}),
  \urlprefix\url{https://link.aps.org/doi/10.1103/PhysRevLett.108.227205}.

\bibitem[{\citenamefont{Oliva-Leyva and Naumis}(2015)}]{non-uniform-strain}
\bibinfo{author}{\bibfnamefont{M.}~\bibnamefont{Oliva-Leyva}} \bibnamefont{and}
  \bibinfo{author}{\bibfnamefont{G.~G.} \bibnamefont{Naumis}},
  \bibinfo{journal}{Physics Letters A} \textbf{\bibinfo{volume}{379}},
  \bibinfo{pages}{2645 } (\bibinfo{year}{2015}), ISSN
  \bibinfo{issn}{0375-9601},
  \urlprefix\url{http://www.sciencedirect.com/science/article/pii/S0375960115005149}.

\bibitem[{\citenamefont{Masir et~al.}(2013)\citenamefont{Masir, Moldovan, and
  Peeters}}]{Peeters-revisited}
\bibinfo{author}{\bibfnamefont{M.~R.} \bibnamefont{Masir}},
  \bibinfo{author}{\bibfnamefont{D.}~\bibnamefont{Moldovan}}, \bibnamefont{and}
  \bibinfo{author}{\bibfnamefont{F.}~\bibnamefont{Peeters}},
  \bibinfo{journal}{Solid State Communications}
  \textbf{\bibinfo{volume}{175-176}}, \bibinfo{pages}{76 }
  (\bibinfo{year}{2013}), ISSN \bibinfo{issn}{0038-1098},
  \bibinfo{note}{special Issue: Graphene V: Recent Advances in Studies of
  Graphene and Graphene analogues},
  \urlprefix\url{http://www.sciencedirect.com/science/article/pii/S0038109813001555}.

\bibitem[{\citenamefont{{Zhai} and {Sandler}}(2018)}]{Valley-filter}
\bibinfo{author}{\bibfnamefont{D.}~\bibnamefont{{Zhai}}} \bibnamefont{and}
  \bibinfo{author}{\bibfnamefont{N.}~\bibnamefont{{Sandler}}},
  \bibinfo{journal}{ArXiv e-prints}  (\bibinfo{year}{2018}),
  \eprint{1806.11251}.

\bibitem[{\citenamefont{Settnes
  et~al.}(2016{\natexlab{a}})\citenamefont{Settnes, Power, Brandbyge, and
  Jauho}}]{Valley-filter2}
\bibinfo{author}{\bibfnamefont{M.}~\bibnamefont{Settnes}},
  \bibinfo{author}{\bibfnamefont{S.~R.} \bibnamefont{Power}},
  \bibinfo{author}{\bibfnamefont{M.}~\bibnamefont{Brandbyge}},
  \bibnamefont{and} \bibinfo{author}{\bibfnamefont{A.-P.} \bibnamefont{Jauho}},
  \bibinfo{journal}{Phys. Rev. Lett.} \textbf{\bibinfo{volume}{117}},
  \bibinfo{pages}{276801} (\bibinfo{year}{2016}{\natexlab{a}}),
  \urlprefix\url{https://link.aps.org/doi/10.1103/PhysRevLett.117.276801}.

\bibitem[{\citenamefont{Tohid and Arash}(2017)}]{Valley-filter3}
\bibinfo{author}{\bibfnamefont{F.}~\bibnamefont{Tohid}} \bibnamefont{and}
  \bibinfo{author}{\bibfnamefont{P.}~\bibnamefont{Arash}},
  \bibinfo{journal}{Scientific Reports} \textbf{\bibinfo{volume}{7}},
  \bibinfo{pages}{17878} (\bibinfo{year}{2017}), ISSN
  \bibinfo{issn}{2045-2322}.

\bibitem[{\citenamefont{Yao et~al.}(2015)\citenamefont{Yao, Liu, Zhu, and
  Zheng}}]{Valley-filter4}
\bibinfo{author}{\bibfnamefont{H.-B.} \bibnamefont{Yao}},
  \bibinfo{author}{\bibfnamefont{Z.}~\bibnamefont{Liu}},
  \bibinfo{author}{\bibfnamefont{M.-F.} \bibnamefont{Zhu}}, \bibnamefont{and}
  \bibinfo{author}{\bibfnamefont{Y.-S.} \bibnamefont{Zheng}},
  \bibinfo{journal}{EPL (Europhysics Letters)} \textbf{\bibinfo{volume}{109}},
  \bibinfo{pages}{37010} (\bibinfo{year}{2015}),
  \urlprefix\url{http://stacks.iop.org/0295-5075/109/i=3/a=37010}.

\bibitem[{\citenamefont{Tony and F.}(2010)}]{Valley-filter5}
\bibinfo{author}{\bibfnamefont{L.}~\bibnamefont{Tony}} \bibnamefont{and}
  \bibinfo{author}{\bibfnamefont{G.}~\bibnamefont{F.}}, \bibinfo{journal}{Nano
  Letters} \textbf{\bibinfo{volume}{10}}, \bibinfo{pages}{3551}
  (\bibinfo{year}{2010}), ISSN \bibinfo{issn}{1530-6984}, \bibinfo{note}{doi:
  10.1021/nl1018063}.

\bibitem[{\citenamefont{Fujita et~al.}(2010)\citenamefont{Fujita, Jalil, and
  Tan}}]{Valley-filter6}
\bibinfo{author}{\bibfnamefont{T.}~\bibnamefont{Fujita}},
  \bibinfo{author}{\bibfnamefont{M.~B.~A.} \bibnamefont{Jalil}},
  \bibnamefont{and} \bibinfo{author}{\bibfnamefont{S.~G.} \bibnamefont{Tan}},
  \bibinfo{journal}{Applied Physics Letters} \textbf{\bibinfo{volume}{97}},
  \bibinfo{pages}{043508} (\bibinfo{year}{2010}),
  \eprint{https://doi.org/10.1063/1.3473725},
  \urlprefix\url{https://doi.org/10.1063/1.3473725}.

\bibitem[{\citenamefont{Zhai et~al.}(2011)\citenamefont{Zhai, Ma, and
  Zhang}}]{Valley-filter7}
\bibinfo{author}{\bibfnamefont{F.}~\bibnamefont{Zhai}},
  \bibinfo{author}{\bibfnamefont{Y.}~\bibnamefont{Ma}}, \bibnamefont{and}
  \bibinfo{author}{\bibfnamefont{Y.-T.} \bibnamefont{Zhang}},
  \bibinfo{journal}{Journal of Physics: Condensed Matter}
  \textbf{\bibinfo{volume}{23}}, \bibinfo{pages}{385302}
  (\bibinfo{year}{2011}),
  \urlprefix\url{http://stacks.iop.org/0953-8984/23/i=38/a=385302}.

\bibitem[{\citenamefont{Zhai et~al.}(2010)\citenamefont{Zhai, Zhao, Chang, and
  Xu}}]{Valley-filter8}
\bibinfo{author}{\bibfnamefont{F.}~\bibnamefont{Zhai}},
  \bibinfo{author}{\bibfnamefont{X.}~\bibnamefont{Zhao}},
  \bibinfo{author}{\bibfnamefont{K.}~\bibnamefont{Chang}}, \bibnamefont{and}
  \bibinfo{author}{\bibfnamefont{H.~Q.} \bibnamefont{Xu}},
  \bibinfo{journal}{Phys. Rev. B} \textbf{\bibinfo{volume}{82}},
  \bibinfo{pages}{115442} (\bibinfo{year}{2010}),
  \urlprefix\url{https://link.aps.org/doi/10.1103/PhysRevB.82.115442}.

\bibitem[{\citenamefont{Chaves et~al.}(2010)\citenamefont{Chaves, Covaci,
  Rakhimov, Farias, and Peeters}}]{Valley-filter9}
\bibinfo{author}{\bibfnamefont{A.}~\bibnamefont{Chaves}},
  \bibinfo{author}{\bibfnamefont{L.}~\bibnamefont{Covaci}},
  \bibinfo{author}{\bibfnamefont{K.~Y.} \bibnamefont{Rakhimov}},
  \bibinfo{author}{\bibfnamefont{G.~A.} \bibnamefont{Farias}},
  \bibnamefont{and} \bibinfo{author}{\bibfnamefont{F.~M.}
  \bibnamefont{Peeters}}, \bibinfo{journal}{Phys. Rev. B}
  \textbf{\bibinfo{volume}{82}}, \bibinfo{pages}{205430}
  (\bibinfo{year}{2010}),
  \urlprefix\url{https://link.aps.org/doi/10.1103/PhysRevB.82.205430}.

\bibitem[{\citenamefont{Jones and Pereira}(2014)}]{Designing}
\bibinfo{author}{\bibfnamefont{G.~W.} \bibnamefont{Jones}} \bibnamefont{and}
  \bibinfo{author}{\bibfnamefont{V.~M.} \bibnamefont{Pereira}},
  \bibinfo{journal}{New Journal of Physics} \textbf{\bibinfo{volume}{16}},
  \bibinfo{pages}{093044} (\bibinfo{year}{2014}),
  \urlprefix\url{http://stacks.iop.org/1367-2630/16/i=9/a=093044}.

\bibitem[{\citenamefont{Klimov et~al.}(2012)\citenamefont{Klimov, Jung, Zhu,
  Li, Wright, Solares, Newell, Zhitenev, and Stroscio}}]{Klimov1557}
\bibinfo{author}{\bibfnamefont{N.~N.} \bibnamefont{Klimov}},
  \bibinfo{author}{\bibfnamefont{S.}~\bibnamefont{Jung}},
  \bibinfo{author}{\bibfnamefont{S.}~\bibnamefont{Zhu}},
  \bibinfo{author}{\bibfnamefont{T.}~\bibnamefont{Li}},
  \bibinfo{author}{\bibfnamefont{C.~A.} \bibnamefont{Wright}},
  \bibinfo{author}{\bibfnamefont{S.~D.} \bibnamefont{Solares}},
  \bibinfo{author}{\bibfnamefont{D.~B.} \bibnamefont{Newell}},
  \bibinfo{author}{\bibfnamefont{N.~B.} \bibnamefont{Zhitenev}},
  \bibnamefont{and} \bibinfo{author}{\bibfnamefont{J.~A.}
  \bibnamefont{Stroscio}}, \bibinfo{journal}{Science}
  \textbf{\bibinfo{volume}{336}}, \bibinfo{pages}{1557} (\bibinfo{year}{2012}),
  ISSN \bibinfo{issn}{0036-8075},
  \eprint{http://science.sciencemag.org/content/336/6088/1557.full.pdf},
  \urlprefix\url{http://science.sciencemag.org/content/336/6088/1557}.

\bibitem[{\citenamefont{Levy et~al.}(2010)\citenamefont{Levy, Burke, Meaker,
  Panlasigui, Zettl, Guinea, Neto, and Crommie}}]{Science-NB}
\bibinfo{author}{\bibfnamefont{N.}~\bibnamefont{Levy}},
  \bibinfo{author}{\bibfnamefont{S.~A.} \bibnamefont{Burke}},
  \bibinfo{author}{\bibfnamefont{K.~L.} \bibnamefont{Meaker}},
  \bibinfo{author}{\bibfnamefont{M.}~\bibnamefont{Panlasigui}},
  \bibinfo{author}{\bibfnamefont{A.}~\bibnamefont{Zettl}},
  \bibinfo{author}{\bibfnamefont{F.}~\bibnamefont{Guinea}},
  \bibinfo{author}{\bibfnamefont{A.~H.~C.} \bibnamefont{Neto}},
  \bibnamefont{and} \bibinfo{author}{\bibfnamefont{M.~F.}
  \bibnamefont{Crommie}}, \bibinfo{journal}{Science}
  \textbf{\bibinfo{volume}{329}}, \bibinfo{pages}{544} (\bibinfo{year}{2010}),
  ISSN \bibinfo{issn}{0036-8075},
  \eprint{http://science.sciencemag.org/content/329/5991/544.full.pdf},
  \urlprefix\url{http://science.sciencemag.org/content/329/5991/544}.

\bibitem[{\citenamefont{Luican et~al.}(2011)\citenamefont{Luican, Li, and
  Andrei}}]{Quantized}
\bibinfo{author}{\bibfnamefont{A.}~\bibnamefont{Luican}},
  \bibinfo{author}{\bibfnamefont{G.}~\bibnamefont{Li}}, \bibnamefont{and}
  \bibinfo{author}{\bibfnamefont{E.~Y.} \bibnamefont{Andrei}},
  \bibinfo{journal}{Phys. Rev. B} \textbf{\bibinfo{volume}{83}},
  \bibinfo{pages}{041405} (\bibinfo{year}{2011}),
  \urlprefix\url{https://link.aps.org/doi/10.1103/PhysRevB.83.041405}.

\bibitem[{\citenamefont{Meng et~al.}(2013)\citenamefont{Meng, He, Zheng, Liu,
  Yan, Yan, Chu, Bai, Dou, Zhang et~al.}}]{PhysRevB.87.205405}
\bibinfo{author}{\bibfnamefont{L.}~\bibnamefont{Meng}},
  \bibinfo{author}{\bibfnamefont{W.-Y.} \bibnamefont{He}},
  \bibinfo{author}{\bibfnamefont{H.}~\bibnamefont{Zheng}},
  \bibinfo{author}{\bibfnamefont{M.}~\bibnamefont{Liu}},
  \bibinfo{author}{\bibfnamefont{H.}~\bibnamefont{Yan}},
  \bibinfo{author}{\bibfnamefont{W.}~\bibnamefont{Yan}},
  \bibinfo{author}{\bibfnamefont{Z.-D.} \bibnamefont{Chu}},
  \bibinfo{author}{\bibfnamefont{K.}~\bibnamefont{Bai}},
  \bibinfo{author}{\bibfnamefont{R.-F.} \bibnamefont{Dou}},
  \bibinfo{author}{\bibfnamefont{Y.}~\bibnamefont{Zhang}},
  \bibnamefont{et~al.}, \bibinfo{journal}{Phys. Rev. B}
  \textbf{\bibinfo{volume}{87}}, \bibinfo{pages}{205405}
  (\bibinfo{year}{2013}),
  \urlprefix\url{https://link.aps.org/doi/10.1103/PhysRevB.87.205405}.

\bibitem[{\citenamefont{Li et~al.}(2015)\citenamefont{Li, Bai, Yin, Qiao, Wang,
  and He}}]{Landau-quantization2}
\bibinfo{author}{\bibfnamefont{S.-Y.} \bibnamefont{Li}},
  \bibinfo{author}{\bibfnamefont{K.-K.} \bibnamefont{Bai}},
  \bibinfo{author}{\bibfnamefont{L.-J.} \bibnamefont{Yin}},
  \bibinfo{author}{\bibfnamefont{J.-B.} \bibnamefont{Qiao}},
  \bibinfo{author}{\bibfnamefont{W.-X.} \bibnamefont{Wang}}, \bibnamefont{and}
  \bibinfo{author}{\bibfnamefont{L.}~\bibnamefont{He}}, \bibinfo{journal}{Phys.
  Rev. B} \textbf{\bibinfo{volume}{92}}, \bibinfo{pages}{245302}
  (\bibinfo{year}{2015}),
  \urlprefix\url{https://link.aps.org/doi/10.1103/PhysRevB.92.245302}.

\bibitem[{\citenamefont{Yan et~al.}(2012)\citenamefont{Yan, Sun, He, Nie, and
  Chan}}]{77K-ridge}
\bibinfo{author}{\bibfnamefont{H.}~\bibnamefont{Yan}},
  \bibinfo{author}{\bibfnamefont{Y.}~\bibnamefont{Sun}},
  \bibinfo{author}{\bibfnamefont{L.}~\bibnamefont{He}},
  \bibinfo{author}{\bibfnamefont{J.-C.} \bibnamefont{Nie}}, \bibnamefont{and}
  \bibinfo{author}{\bibfnamefont{M.~H.~W.} \bibnamefont{Chan}},
  \bibinfo{journal}{Phys. Rev. B} \textbf{\bibinfo{volume}{85}},
  \bibinfo{pages}{035422} (\bibinfo{year}{2012}),
  \urlprefix\url{https://link.aps.org/doi/10.1103/PhysRevB.85.035422}.

\bibitem[{\citenamefont{Ericksen}(2008)}]{CB-Rule}
\bibinfo{author}{\bibfnamefont{J.}~\bibnamefont{Ericksen}},
  \bibinfo{journal}{Mathematics and Mechanics of Solids}
  \textbf{\bibinfo{volume}{13}}, \bibinfo{pages}{199} (\bibinfo{year}{2008}),
  \eprint{https://doi.org/10.1177/1081286507086898},
  \urlprefix\url{https://doi.org/10.1177/1081286507086898}.

\bibitem[{\citenamefont{Wehling
  et~al.}(2008{\natexlab{a}})\citenamefont{Wehling, Balatsky, Tsvelik,
  Katsnelson, and Lichtenstein}}]{midgap}
\bibinfo{author}{\bibfnamefont{T.~O.} \bibnamefont{Wehling}},
  \bibinfo{author}{\bibfnamefont{A.~V.} \bibnamefont{Balatsky}},
  \bibinfo{author}{\bibfnamefont{A.~M.} \bibnamefont{Tsvelik}},
  \bibinfo{author}{\bibfnamefont{M.~I.} \bibnamefont{Katsnelson}},
  \bibnamefont{and} \bibinfo{author}{\bibfnamefont{A.~I.}
  \bibnamefont{Lichtenstein}}, \bibinfo{journal}{EPL (Europhysics Letters)}
  \textbf{\bibinfo{volume}{84}}, \bibinfo{pages}{17003}
  (\bibinfo{year}{2008}{\natexlab{a}}),
  \urlprefix\url{http://stacks.iop.org/0295-5075/84/i=1/a=17003}.

\bibitem[{\citenamefont{Lin et~al.}(2015)\citenamefont{Lin, Chang, Shyu, Lu,
  and Lin}}]{Carbon}
\bibinfo{author}{\bibfnamefont{S.-Y.} \bibnamefont{Lin}},
  \bibinfo{author}{\bibfnamefont{S.-L.} \bibnamefont{Chang}},
  \bibinfo{author}{\bibfnamefont{F.-L.} \bibnamefont{Shyu}},
  \bibinfo{author}{\bibfnamefont{J.-M.} \bibnamefont{Lu}}, \bibnamefont{and}
  \bibinfo{author}{\bibfnamefont{M.-F.} \bibnamefont{Lin}},
  \bibinfo{journal}{Carbon} \textbf{\bibinfo{volume}{86}}, \bibinfo{pages}{207
  } (\bibinfo{year}{2015}), ISSN \bibinfo{issn}{0008-6223},
  \urlprefix\url{http://www.sciencedirect.com/science/article/pii/S0008622314012299}.

\bibitem[{\citenamefont{Verbiest et~al.}(2016)\citenamefont{Verbiest, Stampfer,
  Huber, Andersen, and Reuter}}]{Interplay}
\bibinfo{author}{\bibfnamefont{G.~J.} \bibnamefont{Verbiest}},
  \bibinfo{author}{\bibfnamefont{C.}~\bibnamefont{Stampfer}},
  \bibinfo{author}{\bibfnamefont{S.~E.} \bibnamefont{Huber}},
  \bibinfo{author}{\bibfnamefont{M.}~\bibnamefont{Andersen}}, \bibnamefont{and}
  \bibinfo{author}{\bibfnamefont{K.}~\bibnamefont{Reuter}},
  \bibinfo{journal}{Phys. Rev. B} \textbf{\bibinfo{volume}{93}},
  \bibinfo{pages}{195438} (\bibinfo{year}{2016}),
  \urlprefix\url{https://link.aps.org/doi/10.1103/PhysRevB.93.195438}.

\bibitem[{\citenamefont{Bahamon et~al.}(2015)\citenamefont{Bahamon, Qi, Park,
  Pereira, and Campbell}}]{nanobubbles1}
\bibinfo{author}{\bibfnamefont{D.~A.} \bibnamefont{Bahamon}},
  \bibinfo{author}{\bibfnamefont{Z.}~\bibnamefont{Qi}},
  \bibinfo{author}{\bibfnamefont{H.~S.} \bibnamefont{Park}},
  \bibinfo{author}{\bibfnamefont{V.~M.} \bibnamefont{Pereira}},
  \bibnamefont{and} \bibinfo{author}{\bibfnamefont{D.~K.}
  \bibnamefont{Campbell}}, \bibinfo{journal}{Nanoscale}
  \textbf{\bibinfo{volume}{7}}, \bibinfo{pages}{15300} (\bibinfo{year}{2015}),
  \urlprefix\url{http://dx.doi.org/10.1039/C5NR03393D}.

\bibitem[{\citenamefont{Neek-Amal and Peeters}(2012{\natexlab{a}})}]{Peeters1}
\bibinfo{author}{\bibfnamefont{M.}~\bibnamefont{Neek-Amal}} \bibnamefont{and}
  \bibinfo{author}{\bibfnamefont{F.~M.} \bibnamefont{Peeters}},
  \bibinfo{journal}{Phys. Rev. B} \textbf{\bibinfo{volume}{85}},
  \bibinfo{pages}{195445} (\bibinfo{year}{2012}{\natexlab{a}}),
  \urlprefix\url{https://link.aps.org/doi/10.1103/PhysRevB.85.195445}.

\bibitem[{\citenamefont{Neek-Amal and Peeters}(2012{\natexlab{b}})}]{Peeters2}
\bibinfo{author}{\bibfnamefont{M.}~\bibnamefont{Neek-Amal}} \bibnamefont{and}
  \bibinfo{author}{\bibfnamefont{F.~M.} \bibnamefont{Peeters}},
  \bibinfo{journal}{Phys. Rev. B} \textbf{\bibinfo{volume}{85}},
  \bibinfo{pages}{195446} (\bibinfo{year}{2012}{\natexlab{b}}),
  \urlprefix\url{https://link.aps.org/doi/10.1103/PhysRevB.85.195446}.

\bibitem[{\citenamefont{Neek-Amal et~al.}(2013)\citenamefont{Neek-Amal, Covaci,
  Shakouri, and Peeters}}]{Peeters-triaxial}
\bibinfo{author}{\bibfnamefont{M.}~\bibnamefont{Neek-Amal}},
  \bibinfo{author}{\bibfnamefont{L.}~\bibnamefont{Covaci}},
  \bibinfo{author}{\bibfnamefont{K.}~\bibnamefont{Shakouri}}, \bibnamefont{and}
  \bibinfo{author}{\bibfnamefont{F.~M.} \bibnamefont{Peeters}},
  \bibinfo{journal}{Phys. Rev. B} \textbf{\bibinfo{volume}{88}},
  \bibinfo{pages}{115428} (\bibinfo{year}{2013}),
  \urlprefix\url{https://link.aps.org/doi/10.1103/PhysRevB.88.115428}.

\bibitem[{\citenamefont{Qi et~al.}(2014)\citenamefont{Qi, Kitt, Park, Pereira,
  Campbell, and Castro~Neto}}]{MD-study}
\bibinfo{author}{\bibfnamefont{Z.}~\bibnamefont{Qi}},
  \bibinfo{author}{\bibfnamefont{A.~L.} \bibnamefont{Kitt}},
  \bibinfo{author}{\bibfnamefont{H.~S.} \bibnamefont{Park}},
  \bibinfo{author}{\bibfnamefont{V.~M.} \bibnamefont{Pereira}},
  \bibinfo{author}{\bibfnamefont{D.~K.} \bibnamefont{Campbell}},
  \bibnamefont{and} \bibinfo{author}{\bibfnamefont{A.~H.}
  \bibnamefont{Castro~Neto}}, \bibinfo{journal}{Phys. Rev. B}
  \textbf{\bibinfo{volume}{90}}, \bibinfo{pages}{125419}
  (\bibinfo{year}{2014}),
  \urlprefix\url{https://link.aps.org/doi/10.1103/PhysRevB.90.125419}.

\bibitem[{\citenamefont{Guinea et~al.}(2008)\citenamefont{Guinea, Katsnelson,
  and Vozmediano}}]{midgap2}
\bibinfo{author}{\bibfnamefont{F.}~\bibnamefont{Guinea}},
  \bibinfo{author}{\bibfnamefont{M.~I.} \bibnamefont{Katsnelson}},
  \bibnamefont{and} \bibinfo{author}{\bibfnamefont{M.~A.~H.}
  \bibnamefont{Vozmediano}}, \bibinfo{journal}{Phys. Rev. B}
  \textbf{\bibinfo{volume}{77}}, \bibinfo{pages}{075422}
  (\bibinfo{year}{2008}),
  \urlprefix\url{https://link.aps.org/doi/10.1103/PhysRevB.77.075422}.

\bibitem[{\citenamefont{Zhou and Huang}(2008)}]{Relaxation}
\bibinfo{author}{\bibfnamefont{J.}~\bibnamefont{Zhou}} \bibnamefont{and}
  \bibinfo{author}{\bibfnamefont{R.}~\bibnamefont{Huang}},
  \bibinfo{journal}{Journal of the Mechanics and Physics of Solids}
  \textbf{\bibinfo{volume}{56}}, \bibinfo{pages}{1609 } (\bibinfo{year}{2008}),
  ISSN \bibinfo{issn}{0022-5096},
  \urlprefix\url{http://www.sciencedirect.com/science/article/pii/S0022509607001639}.

\bibitem[{\citenamefont{Guinea et~al.}(2009)\citenamefont{Guinea, Katsnelson,
  and Geim}}]{Guinea2009}
\bibinfo{author}{\bibfnamefont{F.}~\bibnamefont{Guinea}},
  \bibinfo{author}{\bibfnamefont{M.~I.} \bibnamefont{Katsnelson}},
  \bibnamefont{and} \bibinfo{author}{\bibfnamefont{A.~K.} \bibnamefont{Geim}},
  \bibinfo{journal}{Nature Physics} \textbf{\bibinfo{volume}{6}},
  \bibinfo{pages}{30 EP } (\bibinfo{year}{2009}),
  \urlprefix\url{http://dx.doi.org/10.1038/nphys1420}.

\bibitem[{\citenamefont{Georgi et~al.}(2017)\citenamefont{Georgi, Nemes-Incze,
  Carrillo-Bastos, Faria, Viola~Kusminskiy, Zhai, Schneider, Subramaniam,
  Mashoff, Freitag et~al.}}]{multi-uni}
\bibinfo{author}{\bibfnamefont{A.}~\bibnamefont{Georgi}},
  \bibinfo{author}{\bibfnamefont{P.}~\bibnamefont{Nemes-Incze}},
  \bibinfo{author}{\bibfnamefont{R.}~\bibnamefont{Carrillo-Bastos}},
  \bibinfo{author}{\bibfnamefont{D.}~\bibnamefont{Faria}},
  \bibinfo{author}{\bibfnamefont{S.}~\bibnamefont{Viola~Kusminskiy}},
  \bibinfo{author}{\bibfnamefont{D.}~\bibnamefont{Zhai}},
  \bibinfo{author}{\bibfnamefont{M.}~\bibnamefont{Schneider}},
  \bibinfo{author}{\bibfnamefont{D.}~\bibnamefont{Subramaniam}},
  \bibinfo{author}{\bibfnamefont{T.}~\bibnamefont{Mashoff}},
  \bibinfo{author}{\bibfnamefont{N.~M.} \bibnamefont{Freitag}},
  \bibnamefont{et~al.}, \bibinfo{journal}{Nano Letters}
  \textbf{\bibinfo{volume}{17}}, \bibinfo{pages}{2240} (\bibinfo{year}{2017}),
  \bibinfo{note}{pMID: 28211276},
  \eprint{https://doi.org/10.1021/acs.nanolett.6b04870},
  \urlprefix\url{https://doi.org/10.1021/acs.nanolett.6b04870}.

\bibitem[{\citenamefont{Settnes
  et~al.}(2016{\natexlab{b}})\citenamefont{Settnes, Power, and
  Jauho}}]{PhysRevB.93.035456}
\bibinfo{author}{\bibfnamefont{M.}~\bibnamefont{Settnes}},
  \bibinfo{author}{\bibfnamefont{S.~R.} \bibnamefont{Power}}, \bibnamefont{and}
  \bibinfo{author}{\bibfnamefont{A.-P.} \bibnamefont{Jauho}},
  \bibinfo{journal}{Phys. Rev. B} \textbf{\bibinfo{volume}{93}},
  \bibinfo{pages}{035456} (\bibinfo{year}{2016}{\natexlab{b}}),
  \urlprefix\url{https://link.aps.org/doi/10.1103/PhysRevB.93.035456}.

\bibitem[{\citenamefont{Gomes et~al.}(2012)\citenamefont{Gomes, Mar, Ko,
  Guinea, and Manoharan}}]{Gomes2012}
\bibinfo{author}{\bibfnamefont{K.~K.} \bibnamefont{Gomes}},
  \bibinfo{author}{\bibfnamefont{W.}~\bibnamefont{Mar}},
  \bibinfo{author}{\bibfnamefont{W.}~\bibnamefont{Ko}},
  \bibinfo{author}{\bibfnamefont{F.}~\bibnamefont{Guinea}}, \bibnamefont{and}
  \bibinfo{author}{\bibfnamefont{H.~C.} \bibnamefont{Manoharan}},
  \bibinfo{journal}{Nature} \textbf{\bibinfo{volume}{483}}, \bibinfo{pages}{306
  EP } (\bibinfo{year}{2012}),
  \urlprefix\url{http://dx.doi.org/10.1038/nature10941}.

\bibitem[{\citenamefont{{Xu Ke} et~al.}(2009)\citenamefont{{Xu Ke}, {Cao
  Peigen}, and {Heath James R.}}}]{SSB4}
\bibinfo{author}{\bibnamefont{{Xu Ke}}}, \bibinfo{author}{\bibnamefont{{Cao
  Peigen}}}, \bibnamefont{and} \bibinfo{author}{\bibnamefont{{Heath James
  R.}}}, \bibinfo{journal}{Nano Letters} \textbf{\bibinfo{volume}{9}},
  \bibinfo{pages}{4446} (\bibinfo{year}{2009}), ISSN \bibinfo{issn}{1530-6984},
  \bibinfo{note}{doi: 10.1021/nl902729p}.

\bibitem[{\citenamefont{{Lu Jiong} et~al.}(2012)\citenamefont{{Lu Jiong}, {Neto
  A.H. Castro}, and {Loh Kian Ping}}}]{SSB5}
\bibinfo{author}{\bibnamefont{{Lu Jiong}}}, \bibinfo{author}{\bibnamefont{{Neto
  A.H. Castro}}}, \bibnamefont{and} \bibinfo{author}{\bibnamefont{{Loh Kian
  Ping}}}, \bibinfo{journal}{Nature Communications}
  \textbf{\bibinfo{volume}{3}}, \bibinfo{pages}{823} (\bibinfo{year}{2012}),
  \urlprefix\url{https://www.nature.com/articles/ncomms1818\#supplementary-information}.

\bibitem[{\citenamefont{Sun et~al.}(2009)\citenamefont{Sun, Jia, Xue, and
  Li}}]{SSB3}
\bibinfo{author}{\bibfnamefont{G.~F.} \bibnamefont{Sun}},
  \bibinfo{author}{\bibfnamefont{J.~F.} \bibnamefont{Jia}},
  \bibinfo{author}{\bibfnamefont{Q.~K.} \bibnamefont{Xue}}, \bibnamefont{and}
  \bibinfo{author}{\bibfnamefont{L.}~\bibnamefont{Li}},
  \bibinfo{journal}{Nanotechnology} \textbf{\bibinfo{volume}{20}},
  \bibinfo{pages}{355701} (\bibinfo{year}{2009}),
  \urlprefix\url{http://stacks.iop.org/0957-4484/20/i=35/a=355701}.

\bibitem[{\citenamefont{Wakker et~al.}(2011)\citenamefont{Wakker, Tiwari, and
  Blaauboer}}]{currents2}
\bibinfo{author}{\bibfnamefont{G.~M.~M.} \bibnamefont{Wakker}},
  \bibinfo{author}{\bibfnamefont{R.~P.} \bibnamefont{Tiwari}},
  \bibnamefont{and}
  \bibinfo{author}{\bibfnamefont{M.}~\bibnamefont{Blaauboer}},
  \bibinfo{journal}{Phys. Rev. B} \textbf{\bibinfo{volume}{84}},
  \bibinfo{pages}{195427} (\bibinfo{year}{2011}),
  \urlprefix\url{https://link.aps.org/doi/10.1103/PhysRevB.84.195427}.

\bibitem[{\citenamefont{Carrillo-Bastos
  et~al.}(2014)\citenamefont{Carrillo-Bastos, Faria, Latg\'e, Mireles, and
  Sandler}}]{PhysRevB.90.041411}
\bibinfo{author}{\bibfnamefont{R.}~\bibnamefont{Carrillo-Bastos}},
  \bibinfo{author}{\bibfnamefont{D.}~\bibnamefont{Faria}},
  \bibinfo{author}{\bibfnamefont{A.}~\bibnamefont{Latg\'e}},
  \bibinfo{author}{\bibfnamefont{F.}~\bibnamefont{Mireles}}, \bibnamefont{and}
  \bibinfo{author}{\bibfnamefont{N.}~\bibnamefont{Sandler}},
  \bibinfo{journal}{Phys. Rev. B} \textbf{\bibinfo{volume}{90}},
  \bibinfo{pages}{041411} (\bibinfo{year}{2014}),
  \urlprefix\url{https://link.aps.org/doi/10.1103/PhysRevB.90.041411}.

\bibitem[{\citenamefont{Faria et~al.}(2013)\citenamefont{Faria, Latg\'e, Ulloa,
  and Sandler}}]{currents3}
\bibinfo{author}{\bibfnamefont{D.}~\bibnamefont{Faria}},
  \bibinfo{author}{\bibfnamefont{A.}~\bibnamefont{Latg\'e}},
  \bibinfo{author}{\bibfnamefont{S.~E.} \bibnamefont{Ulloa}}, \bibnamefont{and}
  \bibinfo{author}{\bibfnamefont{N.}~\bibnamefont{Sandler}},
  \bibinfo{journal}{Phys. Rev. B} \textbf{\bibinfo{volume}{87}},
  \bibinfo{pages}{241403} (\bibinfo{year}{2013}),
  \urlprefix\url{https://link.aps.org/doi/10.1103/PhysRevB.87.241403}.

\bibitem[{\citenamefont{Schneider et~al.}(2015)\citenamefont{Schneider, Faria,
  Viola~Kusminskiy, and Sandler}}]{Local}
\bibinfo{author}{\bibfnamefont{M.}~\bibnamefont{Schneider}},
  \bibinfo{author}{\bibfnamefont{D.}~\bibnamefont{Faria}},
  \bibinfo{author}{\bibfnamefont{S.}~\bibnamefont{Viola~Kusminskiy}},
  \bibnamefont{and} \bibinfo{author}{\bibfnamefont{N.}~\bibnamefont{Sandler}},
  \bibinfo{journal}{Phys. Rev. B} \textbf{\bibinfo{volume}{91}},
  \bibinfo{pages}{161407} (\bibinfo{year}{2015}),
  \urlprefix\url{https://link.aps.org/doi/10.1103/PhysRevB.91.161407}.

\bibitem[{\citenamefont{Moldovan et~al.}(2013)\citenamefont{Moldovan,
  Ramezani~Masir, and Peeters}}]{Electronic}
\bibinfo{author}{\bibfnamefont{D.}~\bibnamefont{Moldovan}},
  \bibinfo{author}{\bibfnamefont{M.}~\bibnamefont{Ramezani~Masir}},
  \bibnamefont{and} \bibinfo{author}{\bibfnamefont{F.~M.}
  \bibnamefont{Peeters}}, \bibinfo{journal}{Phys. Rev. B}
  \textbf{\bibinfo{volume}{88}}, \bibinfo{pages}{035446}
  (\bibinfo{year}{2013}),
  \urlprefix\url{https://link.aps.org/doi/10.1103/PhysRevB.88.035446}.

\bibitem[{\citenamefont{Ray et~al.}(2016{\natexlab{a}})\citenamefont{Ray, Rost,
  Weckbecker, Vogl, Sharma, Gupta, Pankratov, and Shallcross}}]{M}
\bibinfo{author}{\bibfnamefont{N.}~\bibnamefont{Ray}},
  \bibinfo{author}{\bibfnamefont{F.}~\bibnamefont{Rost}},
  \bibinfo{author}{\bibfnamefont{D.}~\bibnamefont{Weckbecker}},
  \bibinfo{author}{\bibfnamefont{M.}~\bibnamefont{Vogl}},
  \bibinfo{author}{\bibfnamefont{S.}~\bibnamefont{Sharma}},
  \bibinfo{author}{\bibfnamefont{R.}~\bibnamefont{Gupta}},
  \bibinfo{author}{\bibfnamefont{O.}~\bibnamefont{Pankratov}},
  \bibnamefont{and}
  \bibinfo{author}{\bibfnamefont{S.}~\bibnamefont{Shallcross}},
  \bibinfo{journal}{arXiv:1607.00920}  (\bibinfo{year}{2016}{\natexlab{a}}).

\bibitem[{\citenamefont{Jang et~al.}(2014)\citenamefont{Jang, Kim, Shin, Wang,
  Jang, Kim, Lee, Kim, Song, and Kahng}}]{JANG2014139}
\bibinfo{author}{\bibfnamefont{W.-J.} \bibnamefont{Jang}},
  \bibinfo{author}{\bibfnamefont{H.}~\bibnamefont{Kim}},
  \bibinfo{author}{\bibfnamefont{Y.-R.} \bibnamefont{Shin}},
  \bibinfo{author}{\bibfnamefont{M.}~\bibnamefont{Wang}},
  \bibinfo{author}{\bibfnamefont{S.~K.} \bibnamefont{Jang}},
  \bibinfo{author}{\bibfnamefont{M.}~\bibnamefont{Kim}},
  \bibinfo{author}{\bibfnamefont{S.}~\bibnamefont{Lee}},
  \bibinfo{author}{\bibfnamefont{S.-W.} \bibnamefont{Kim}},
  \bibinfo{author}{\bibfnamefont{Y.~J.} \bibnamefont{Song}}, \bibnamefont{and}
  \bibinfo{author}{\bibfnamefont{S.-J.} \bibnamefont{Kahng}},
  \bibinfo{journal}{Carbon} \textbf{\bibinfo{volume}{74}}, \bibinfo{pages}{139
  } (\bibinfo{year}{2014}), ISSN \bibinfo{issn}{0008-6223},
  \urlprefix\url{http://www.sciencedirect.com/science/article/pii/S0008622314002619}.

\bibitem[{\citenamefont{Midtvedt et~al.}(2016)\citenamefont{Midtvedt,
  Lewenkopf, and Croy}}]{non-uniform}
\bibinfo{author}{\bibfnamefont{D.}~\bibnamefont{Midtvedt}},
  \bibinfo{author}{\bibfnamefont{C.~H.} \bibnamefont{Lewenkopf}},
  \bibnamefont{and} \bibinfo{author}{\bibfnamefont{A.}~\bibnamefont{Croy}},
  \bibinfo{journal}{2D Materials} \textbf{\bibinfo{volume}{3}},
  \bibinfo{pages}{011005} (\bibinfo{year}{2016}),
  \urlprefix\url{http://stacks.iop.org/2053-1583/3/i=1/a=011005}.

\bibitem[{\citenamefont{Linnik}(2012)}]{lin12}
\bibinfo{author}{\bibfnamefont{T.~L.} \bibnamefont{Linnik}},
  \bibinfo{journal}{Journal of Physics: Condensed Matter}
  \textbf{\bibinfo{volume}{24}}, \bibinfo{pages}{205302}
  (\bibinfo{year}{2012}),
  \urlprefix\url{http://stacks.iop.org/0953-8984/24/i=20/a=205302}.

\bibitem[{\citenamefont{{Kisslinger Ferdinand}
  et~al.}(2015)\citenamefont{{Kisslinger Ferdinand}, {Ott Christian}, {Heide
  Christian}, {Kampert Erik}, {Butz Benjamin}, {Spiecker Erdmann}, {Shallcross
  Sam}, and {Weber Heiko B.}}}]{kiss15}
\bibinfo{author}{\bibnamefont{{Kisslinger Ferdinand}}},
  \bibinfo{author}{\bibnamefont{{Ott Christian}}},
  \bibinfo{author}{\bibnamefont{{Heide Christian}}},
  \bibinfo{author}{\bibnamefont{{Kampert Erik}}},
  \bibinfo{author}{\bibnamefont{{Butz Benjamin}}},
  \bibinfo{author}{\bibnamefont{{Spiecker Erdmann}}},
  \bibinfo{author}{\bibnamefont{{Shallcross Sam}}}, \bibnamefont{and}
  \bibinfo{author}{\bibnamefont{{Weber Heiko B.}}}, \bibinfo{journal}{Nature
  Physics} \textbf{\bibinfo{volume}{11}}, \bibinfo{pages}{650}
  (\bibinfo{year}{2015}),
  \urlprefix\url{https://www.nature.com/articles/nphys3368\#supplementary-information}.

\bibitem[{\citenamefont{Vogl et~al.}(2017)\citenamefont{Vogl, Pankratov, and
  Shallcross}}]{vogl16}
\bibinfo{author}{\bibfnamefont{M.}~\bibnamefont{Vogl}},
  \bibinfo{author}{\bibfnamefont{O.}~\bibnamefont{Pankratov}},
  \bibnamefont{and}
  \bibinfo{author}{\bibfnamefont{S.}~\bibnamefont{Shallcross}},
  \bibinfo{journal}{Phys. Rev. B} \textbf{\bibinfo{volume}{96}},
  \bibinfo{pages}{035442} (\bibinfo{year}{2017}),
  \urlprefix\url{https://link.aps.org/doi/10.1103/PhysRevB.96.035442}.

\bibitem[{\citenamefont{Ray et~al.}(2016{\natexlab{b}})\citenamefont{Ray,
  Fleischmann, Weckbecker, Sharma, Pankratov, and Shallcross}}]{ray16}
\bibinfo{author}{\bibfnamefont{N.}~\bibnamefont{Ray}},
  \bibinfo{author}{\bibfnamefont{M.}~\bibnamefont{Fleischmann}},
  \bibinfo{author}{\bibfnamefont{D.}~\bibnamefont{Weckbecker}},
  \bibinfo{author}{\bibfnamefont{S.}~\bibnamefont{Sharma}},
  \bibinfo{author}{\bibfnamefont{O.}~\bibnamefont{Pankratov}},
  \bibnamefont{and}
  \bibinfo{author}{\bibfnamefont{S.}~\bibnamefont{Shallcross}},
  \bibinfo{journal}{Phys. Rev. B} \textbf{\bibinfo{volume}{94}},
  \bibinfo{pages}{245403} (\bibinfo{year}{2016}{\natexlab{b}}),
  \urlprefix\url{https://link.aps.org/doi/10.1103/PhysRevB.94.245403}.

\bibitem[{\citenamefont{Ma\~nes}(2007{\natexlab{b}})}]{PhysRevB.76.045430}
\bibinfo{author}{\bibfnamefont{J.~L.} \bibnamefont{Ma\~nes}},
  \bibinfo{journal}{Phys. Rev. B} \textbf{\bibinfo{volume}{76}},
  \bibinfo{pages}{045430} (\bibinfo{year}{2007}{\natexlab{b}}),
  \urlprefix\url{https://link.aps.org/doi/10.1103/PhysRevB.76.045430}.

\bibitem[{\citenamefont{Oliva-Leyva and Naumis}(2016)}]{Optical-cond}
\bibinfo{author}{\bibfnamefont{M.}~\bibnamefont{Oliva-Leyva}} \bibnamefont{and}
  \bibinfo{author}{\bibfnamefont{G.~G.} \bibnamefont{Naumis}},
  \bibinfo{journal}{Phys. Rev. B} \textbf{\bibinfo{volume}{93}},
  \bibinfo{pages}{035439} (\bibinfo{year}{2016}),
  \urlprefix\url{https://link.aps.org/doi/10.1103/PhysRevB.93.035439}.

\bibitem[{\citenamefont{{Carrillo-Bastos}
  et~al.}(2014)\citenamefont{{Carrillo-Bastos}, {Faria}, {Latg{\'e}},
  {Mireles}, and {Sandler}}}]{Gaussian}
\bibinfo{author}{\bibfnamefont{R.}~\bibnamefont{{Carrillo-Bastos}}},
  \bibinfo{author}{\bibfnamefont{D.}~\bibnamefont{{Faria}}},
  \bibinfo{author}{\bibfnamefont{A.}~\bibnamefont{{Latg{\'e}}}},
  \bibinfo{author}{\bibfnamefont{F.}~\bibnamefont{{Mireles}}},
  \bibnamefont{and}
  \bibinfo{author}{\bibfnamefont{N.}~\bibnamefont{{Sandler}}},
  \bibinfo{journal}{\prb} \textbf{\bibinfo{volume}{90}}, \bibinfo{eid}{041411}
  (\bibinfo{year}{2014}), \eprint{1405.1962}.

\bibitem[{\citenamefont{Fleischmann et~al.}(2018)\citenamefont{Fleischmann,
  Gupta, Weckbecker, Landgraf, Pankratov, Meded, and Shallcross}}]{shall18}
\bibinfo{author}{\bibfnamefont{M.}~\bibnamefont{Fleischmann}},
  \bibinfo{author}{\bibfnamefont{R.}~\bibnamefont{Gupta}},
  \bibinfo{author}{\bibfnamefont{D.}~\bibnamefont{Weckbecker}},
  \bibinfo{author}{\bibfnamefont{W.}~\bibnamefont{Landgraf}},
  \bibinfo{author}{\bibfnamefont{O.}~\bibnamefont{Pankratov}},
  \bibinfo{author}{\bibfnamefont{V.}~\bibnamefont{Meded}}, \bibnamefont{and}
  \bibinfo{author}{\bibfnamefont{S.}~\bibnamefont{Shallcross}},
  \bibinfo{journal}{Phys. Rev. B} \textbf{\bibinfo{volume}{97}},
  \bibinfo{pages}{205128} (\bibinfo{year}{2018}),
  \urlprefix\url{https://link.aps.org/doi/10.1103/PhysRevB.97.205128}.

\bibitem[{\citenamefont{de~Juan et~al.}(2013)\citenamefont{de~Juan, Ma\~nes,
  and Vozmediano}}]{PhysRevB.87.165131}
\bibinfo{author}{\bibfnamefont{F.}~\bibnamefont{de~Juan}},
  \bibinfo{author}{\bibfnamefont{J.~L.} \bibnamefont{Ma\~nes}},
  \bibnamefont{and} \bibinfo{author}{\bibfnamefont{M.~A.~H.}
  \bibnamefont{Vozmediano}}, \bibinfo{journal}{Phys. Rev. B}
  \textbf{\bibinfo{volume}{87}}, \bibinfo{pages}{165131}
  (\bibinfo{year}{2013}),
  \urlprefix\url{https://link.aps.org/doi/10.1103/PhysRevB.87.165131}.

\bibitem[{\citenamefont{Stegmann and Szpak}(2016)}]{Nikodem}
\bibinfo{author}{\bibfnamefont{T.}~\bibnamefont{Stegmann}} \bibnamefont{and}
  \bibinfo{author}{\bibfnamefont{N.}~\bibnamefont{Szpak}},
  \bibinfo{journal}{New Journal of Physics} \textbf{\bibinfo{volume}{18}},
  \bibinfo{pages}{053016} (\bibinfo{year}{2016}),
  \urlprefix\url{http://stacks.iop.org/1367-2630/18/i=5/a=053016}.

\bibitem[{\citenamefont{{Yang}}(2012)}]{eddy_currents}
\bibinfo{author}{\bibfnamefont{H.-T.} \bibnamefont{{Yang}}},
  \bibinfo{journal}{ArXiv e-prints}  (\bibinfo{year}{2012}),
  \eprint{1210.1727}.

\bibitem[{\citenamefont{Lee et~al.}(2004)\citenamefont{Lee, Souma, Ihm, and
  Chang}}]{1998}
\bibinfo{author}{\bibfnamefont{S.}~\bibnamefont{Lee}},
  \bibinfo{author}{\bibfnamefont{S.}~\bibnamefont{Souma}},
  \bibinfo{author}{\bibfnamefont{G.}~\bibnamefont{Ihm}}, \bibnamefont{and}
  \bibinfo{author}{\bibfnamefont{K.}~\bibnamefont{Chang}},
  \bibinfo{journal}{Physics Reports} \textbf{\bibinfo{volume}{394}},
  \bibinfo{pages}{1 } (\bibinfo{year}{2004}), ISSN \bibinfo{issn}{0370-1573},
  \urlprefix\url{http://www.sciencedirect.com/science/article/pii/S0370157303004654}.

\bibitem[{\citenamefont{Richter and Sieber}(2002)}]{1999}
\bibinfo{author}{\bibfnamefont{K.}~\bibnamefont{Richter}} \bibnamefont{and}
  \bibinfo{author}{\bibfnamefont{M.}~\bibnamefont{Sieber}},
  \bibinfo{journal}{Phys. Rev. Lett.} \textbf{\bibinfo{volume}{89}},
  \bibinfo{pages}{206801} (\bibinfo{year}{2002}),
  \urlprefix\url{https://link.aps.org/doi/10.1103/PhysRevLett.89.206801}.

\bibitem[{\citenamefont{Reijniers and Peeters}(2000)}]{2000}
\bibinfo{author}{\bibfnamefont{J.}~\bibnamefont{Reijniers}} \bibnamefont{and}
  \bibinfo{author}{\bibfnamefont{F.~M.} \bibnamefont{Peeters}},
  \bibinfo{journal}{Journal of Physics: Condensed Matter}
  \textbf{\bibinfo{volume}{12}}, \bibinfo{pages}{9771} (\bibinfo{year}{2000}),
  \urlprefix\url{http://stacks.iop.org/0953-8984/12/i=47/a=305}.

\bibitem[{\citenamefont{Oroszl\'any et~al.}(2008)\citenamefont{Oroszl\'any,
  Rakyta, Korm\'anyos, Lambert, and Cserti}}]{2008}
\bibinfo{author}{\bibfnamefont{L.}~\bibnamefont{Oroszl\'any}},
  \bibinfo{author}{\bibfnamefont{P.}~\bibnamefont{Rakyta}},
  \bibinfo{author}{\bibfnamefont{A.}~\bibnamefont{Korm\'anyos}},
  \bibinfo{author}{\bibfnamefont{C.~J.} \bibnamefont{Lambert}},
  \bibnamefont{and} \bibinfo{author}{\bibfnamefont{J.}~\bibnamefont{Cserti}},
  \bibinfo{journal}{Phys. Rev. B} \textbf{\bibinfo{volume}{77}},
  \bibinfo{pages}{081403} (\bibinfo{year}{2008}),
  \urlprefix\url{https://link.aps.org/doi/10.1103/PhysRevB.77.081403}.

\bibitem[{\citenamefont{Kim et~al.}(2011)\citenamefont{Kim, Blanter, and
  Ahn}}]{2011}
\bibinfo{author}{\bibfnamefont{K.-J.} \bibnamefont{Kim}},
  \bibinfo{author}{\bibfnamefont{Y.~M.} \bibnamefont{Blanter}},
  \bibnamefont{and} \bibinfo{author}{\bibfnamefont{K.-H.} \bibnamefont{Ahn}},
  \bibinfo{journal}{Phys. Rev. B} \textbf{\bibinfo{volume}{84}},
  \bibinfo{pages}{081401} (\bibinfo{year}{2011}),
  \urlprefix\url{https://link.aps.org/doi/10.1103/PhysRevB.84.081401}.

\bibitem[{\citenamefont{Liu et~al.}(2015)\citenamefont{Liu, Tiwari, Brada,
  Bruder, Kusmartsev, and Mele}}]{2015}
\bibinfo{author}{\bibfnamefont{Y.}~\bibnamefont{Liu}},
  \bibinfo{author}{\bibfnamefont{R.~P.} \bibnamefont{Tiwari}},
  \bibinfo{author}{\bibfnamefont{M.}~\bibnamefont{Brada}},
  \bibinfo{author}{\bibfnamefont{C.}~\bibnamefont{Bruder}},
  \bibinfo{author}{\bibfnamefont{F.~V.} \bibnamefont{Kusmartsev}},
  \bibnamefont{and} \bibinfo{author}{\bibfnamefont{E.~J.} \bibnamefont{Mele}},
  \bibinfo{journal}{Phys. Rev. B} \textbf{\bibinfo{volume}{92}},
  \bibinfo{pages}{235438} (\bibinfo{year}{2015}),
  \urlprefix\url{https://link.aps.org/doi/10.1103/PhysRevB.92.235438}.

\bibitem[{\citenamefont{Castro et~al.}(2017)\citenamefont{Castro, Cazalilla,
  and Vozmediano}}]{PhysRevB.96.241405}
\bibinfo{author}{\bibfnamefont{E.~V.} \bibnamefont{Castro}},
  \bibinfo{author}{\bibfnamefont{M.~A.} \bibnamefont{Cazalilla}},
  \bibnamefont{and} \bibinfo{author}{\bibfnamefont{M.~A.~H.}
  \bibnamefont{Vozmediano}}, \bibinfo{journal}{Phys. Rev. B}
  \textbf{\bibinfo{volume}{96}}, \bibinfo{pages}{241405}
  (\bibinfo{year}{2017}),
  \urlprefix\url{https://link.aps.org/doi/10.1103/PhysRevB.96.241405}.

\bibitem[{\citenamefont{Wehling
  et~al.}(2008{\natexlab{b}})\citenamefont{Wehling, Balatsky, Tsvelik,
  Katsnelson, and Lichtenstein}}]{0295-5075-84-1-17003}
\bibinfo{author}{\bibfnamefont{T.~O.} \bibnamefont{Wehling}},
  \bibinfo{author}{\bibfnamefont{A.~V.} \bibnamefont{Balatsky}},
  \bibinfo{author}{\bibfnamefont{A.~M.} \bibnamefont{Tsvelik}},
  \bibinfo{author}{\bibfnamefont{M.~I.} \bibnamefont{Katsnelson}},
  \bibnamefont{and} \bibinfo{author}{\bibfnamefont{A.~I.}
  \bibnamefont{Lichtenstein}}, \bibinfo{journal}{EPL (Europhysics Letters)}
  \textbf{\bibinfo{volume}{84}}, \bibinfo{pages}{17003}
  (\bibinfo{year}{2008}{\natexlab{b}}),
  \urlprefix\url{http://stacks.iop.org/0295-5075/84/i=1/a=17003}.

\bibitem[{\citenamefont{{Oliva-Leyva} and {Wang}}(2018)}]{non-uniform2}
\bibinfo{author}{\bibfnamefont{M.}~\bibnamefont{{Oliva-Leyva}}}
  \bibnamefont{and} \bibinfo{author}{\bibfnamefont{C.}~\bibnamefont{{Wang}}},
  \bibinfo{journal}{ArXiv e-prints}  (\bibinfo{year}{2018}),
  \eprint{1807.02147}.

\end{thebibliography}

\end{document}